\documentclass[aps,pre,reprint,showkeys,floatfix]{revtex4-1}
\usepackage[dvips, xdvi]{graphicx}
\DeclareGraphicsExtensions{.eps}
\usepackage{amsmath}
\usepackage{natbib}
\usepackage{graphics}
\usepackage{amssymb}

\newcommand{\D}{\displaystyle}
\def \Cai {$[{\rm Ca}^{2+}]_i$}
\def \Ca {${\rm Ca}^{2+}$}

\begin{document}

\title{The Myogenic Response in Isolated Rat Cerebrovascular Arteries: Smooth Muscle Cell Model}

\author{Jin Yang}
\email[Electronic address: ]{jinyang2004@gmail.com.} 
\thanks{Current address: CAS-MPG Partner Institute for Computational Biology, Shanghai Institutes for Biological Sciences, 320 Yue Yang Rd., Shanghai 200031, China}
\affiliation{Department of Bioengineering, Rice University, Houston, TX 77005}
\author{John W. Clark, Jr.}
\affiliation{Department of Electrical Engineering, Rice University, Houston, TX 77005}
\author{Robert M. Bryan}
\affiliation{Department of Anesthesiology, Baylor College of Medicine, Houston, TX 77030}
\author{Claudia S. Robertson}
\affiliation{Department of Neurosurgery, Baylor College of Medicine, Houston, TX 77030}

\begin{abstract}
Previous models of the cerebrovascular smooth muscle cell have not addressed the interaction between the electrical, chemical and mechanical components of cell function during the development of active tension. These models are primarily electrical, biochemical or mechanical in their orientation, and do not permit a full exploration of how the smooth muscle responds to electrical or mechanical forcing. To address this issue, we have developed a new model that consists of two major components: electrochemical and chemomechanical subsystem models of the smooth muscle cell. Included in the electrochemical model are models of the electrophysiological behavior of the cell membrane, fluid compartments, \Ca \ release and uptake by the sarcoplasmic reticulum, and cytosolic \Ca \ buffering, particularly by calmodulin. With this subsystem model, we can study the mechanics of the production of intracellular \Ca \ transient in response to membrane voltage clamp pulses. The chemomechanical model includes models of: (a) the chemical kinetics of myosin phosphorylation, and the
formation of phosphorylated (cycling) myosin cross-bridges with actin, as well as, attached (non-cycling) latch-type cross-bridges; and (b) a model of force generation and mechanical coupling to the contractile filaments and their attachments to protein structures and the skeletal framework of the cell. The two subsystem models are tested independently and compared with data. Likewise, the complete (combined) cell model responses to voltage pulse stimulation under isometric and isotonic conditions are calculated and compared with measured single cell length-force and force-velocity data obtained from literature. This integrated cell model provides biophysically based explanations of electrical, chemical and mechanical phenomena in cerebrovascular smooth muscle,  and has considerable utility as an adjunct to laboratory research and experimental design.

\end{abstract}

\keywords{Mechanistic model, intracellular calcium dynamics, myosin phosphorylation, active tension generation}
\maketitle

\section{INTRODUCTION}
Contraction in vascular smooth muscle cells can be initiated by mechanical, electrical, and chemical stimuli. Several different signal transduction pathways activate the contractile mechanism, but all of which first lead to an increase in intracellular \Ca \ concentration \Cai. This increase in \Cai can arise from: (a) \Ca \ influx through voltage-dependent sarcolemmal \Ca \ channels, and/or (b) a release of \Ca \ from internal \Ca \ stores [e.g., from the sarcoplasmic reticulum (SR)]. In the cytosol, free \Ca \ binds to a special \Ca \ binding protein called calmodulin (CM), and the calcium-calmodulin complex (CaCM) activates myosin light chain kinase (MLCK), an enzyme that facilitates the phosphorylation of myosin light chains (MLC) in the presence of ATP. This MLC phosphorylation leads to the cross-bridge formation and cycling of the myosin heads along the actin filaments, generating the active force necessary for muscle contraction. The stressed actin filaments are coupled via a viscoelastic system to the cell membrane, resulting in changes in cell length. Experimentally, the macroscopic contractile properties of a single smooth muscle cell are usually measured in terms of length-force (L:F) and force-velocity (F:V) relationships.

Several models have been developed to investigate the particular aspects of cerebrovascular smooth muscle function, for example, including membrane electrophysiology~\cite{lang:96}, cytosolic calcium regulation~\cite{wong:93}, phosphorylation of myosin~\cite{hai:88a,hai:88b} or mechanical behavior~\cite{per:86}). However, none has sought to integrate these components into a single functional model. In this study, we develop an integrated model of vascular smooth muscle function based on the electrophysiological, biochemical and mechanical data available on isolated smooth muscle cells.

\begin{figure*}
\begin{center}
\includegraphics[scale=0.8]{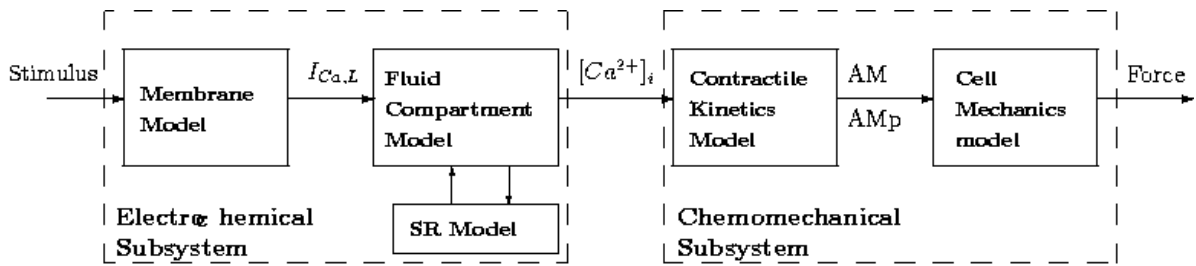}
\caption{\label{fmodel}\textbf{Functional Block Diagram of the vascular smooth muscle cell}. The model consists of two coupled subsystems (electrochemical and chemomechanical). The membrane model describes the electrical response of the cell membrane to stimulation and especially the L-type \Ca \ current ($I_{\rm Ca,L}$), which brings \Ca \ into the cell. The intracellular fluid compartment model is based on material balances of ${\rm Na}^+$, ${\rm K}^+$ and \Ca, whereas the SR model contains a \Ca-induced \Ca-release (CICR) mechanism that releases \Ca \ into the cytosol, and  actively recovers it from the cytosol. The activated calcium-calmodulin (CaCM) drives the phosphorylation of the contractile kinetics model and subsequently active force development by the cell mechanics model. AMp and AM represent phosphorylated and nonphosphorylated states of MLC that forms a crossbridge attachment to the actin filament (See the main text for details).}
\end{center}
\end{figure*}

Figure~\ref{fmodel} shows a functional block diagram of the model for the vascular smooth muscle cell. The whole model is composed of two specific subsystem models. The first is an {\it electrochemical} model that combines a Hodgkin-Huxley (HH)-type model of membrane electrophysiology with models of the intracellular fluid compartment and the sarcoplasmic reticulum compartment. The membrane model contains descriptions of the known ionic membrane currents active in cerebrovascular smooth muscle cells. Material balance equations for cytosolic ionic concentrations of ${\rm Na}^+$, ${\rm K}^+$, and \Ca \ provide the basis of the fluid compartment model under the assumption that the concentrations of these ionic species in the extracellular medium are held at constant levels (see Table~\ref{tabpara1} for numerical values), and that \Ca \ concentration in the cytosol is buffered by the \Ca \ binding
protein calmodulin (CM) and other \Ca \ binding proteins. The material balance for \Ca \ also considers the \Ca \ fluxes entering
and leaving the cell cytosol via the SR and sarcolemmal membranes. The experimental data used to validate the electrophysiological
properties of the membrane and fluid compartment models comes largely from voltage clamp experiments on isolated cells. This
subsystem model thus characterizes the electrochemical behavior of the cell and most importantly the regulation of intracellular \Ca.

The second subsystem model is a {\it chemomechanical} model that has two coupled units: (i) a biochemical model of CaCM-dependent myosin light chain  phosphorylation and attached cross-bridge formation kinetics, and (ii) a mechanical model of force generation and mechanical coupling within the cell. A modified version of the multiple-state kinetic model of myosin phosphorylation originally developed by Hai and Murphy~\cite{hai:88a,hai:88b} is used to represent the dynamics of myosin-actin cross-bridge attachment and detachment, and a mechanical model is used to describe the active force generation by the contractile filaments and the viscoelastic properties of their coupling and attachment to the cell wall. We use data from the length-tension studies of Fay and Warshaw~\cite{dmw:91,dmw:87,fay:83,fay:88b,fay:88a} on isolated smooth muscle cells to help validate this chemomechanical model.

\section{MODEL DEVELOPMENT}

\subsection{Electrochemical model}

\begin{figure}[t]
\begin{center}
\includegraphics[scale=0.22]{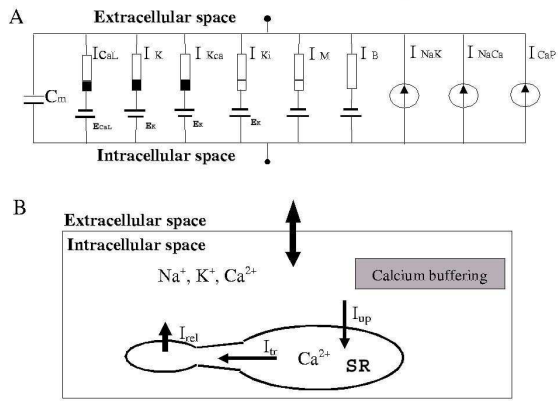}
\caption{\label{fig1} The electrochemical Model. A. Membrane model describing ionic membrane currents and transmembrane potential. Black and white indicates resistance with voltage-dependent nonlinearity and the all white resistor indicates linear element. B. Fluid compartment model describing ionic dynamics, \Ca \ buffering, and \Ca \ handling by sarcoplasmic reticulum (SR).}
\end{center}
\end{figure}

Figure~\ref{fig1}A represents a lumped Hodgkin-Huxley type electrical equivalent circuit of the smooth cell membrane, which consists of a whole-cell membrane capacitance ($C_m$) shunted by a variety of resistive transmembrane channels, as well as, ionic pump and exchanger currents. The Kirchhoff's current law applied to the circuit of Fig.~\ref{fig1}A yields the following differential equation describing changes in the transmembrane potential $V_m$:
\begin{eqnarray}
\notag \frac{dV_m}{dt} & = & -\frac{1}{C_m}(I_{\rm Ca,L}+I_K+I_{\rm K,Ca}+I_{\rm Ki}+ \\
                & & I_{\rm M}+I_{\rm NaCa}+I_{\rm NaK}+I_{\rm CaP}+I_B)
\end{eqnarray}
The individual membrane currents involved in this equation are discussed in the subsequent sections.

To account for temporal and voltage-dependent changes in the ionic concentrations of ${\rm Na}^+$, ${\rm K}^+$ and \Ca \ in the cytosol, a fluid compartment model is also (Fig.~\ref{fig1}B) developed. Cytosolic material balance equations (see Appendix I, Table~\ref{tabmb}) are given for ${\rm Na}^+$, \Ca \ and ${\rm K}^+$. Concentrations for these ions are assumed to be constant in the bulk extracellular bathing medium.  The material balance equation for \Ca \ accounts for \Ca-binding to the cytosolic protein calmodulin and other nonspecific buffering media, \Ca \ uptake and release by the sarcoplasmic reticulum.

\begin{figure}[t]
\begin{center}
  \includegraphics[scale=0.38]{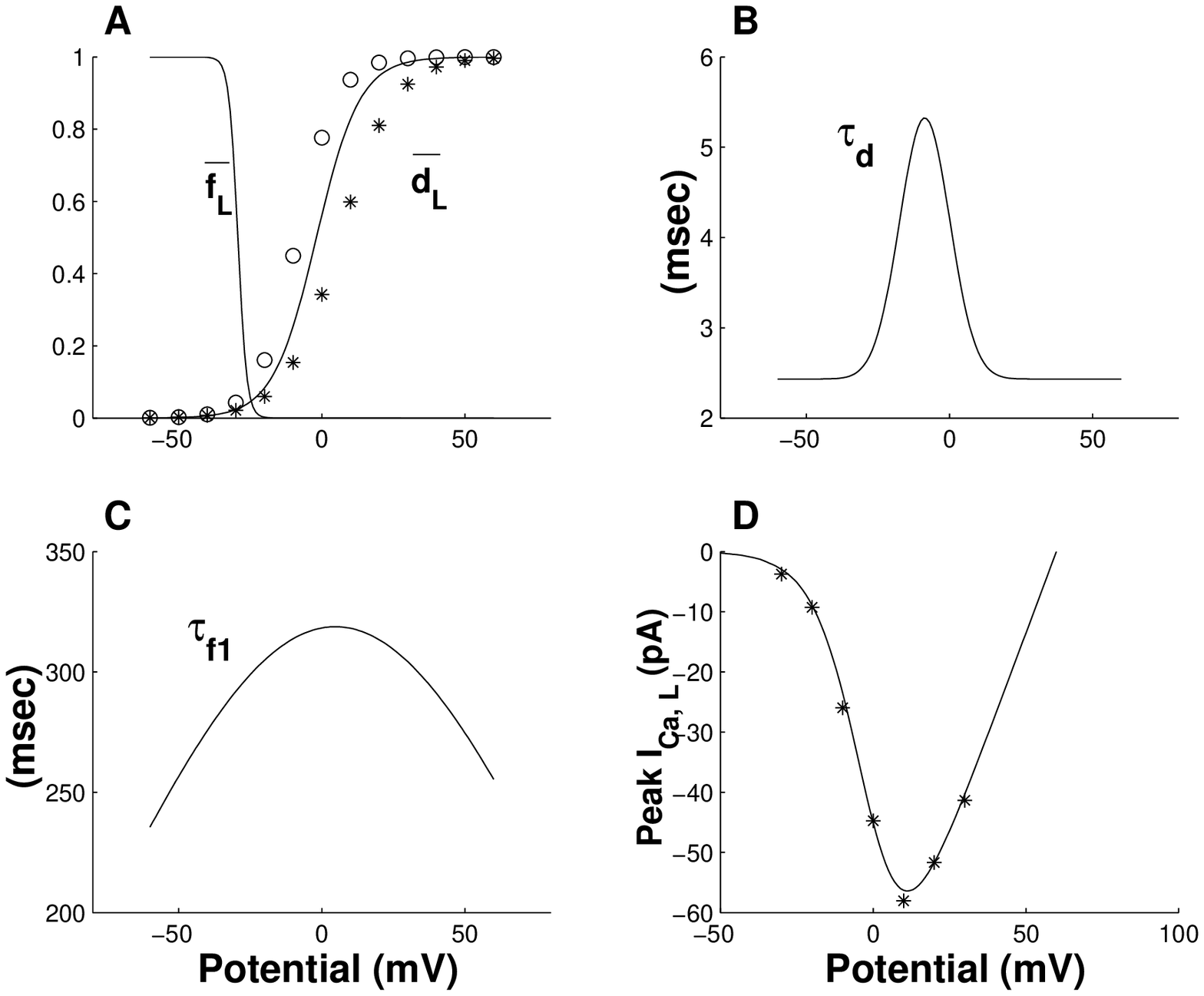}
  \includegraphics[scale=0.38]{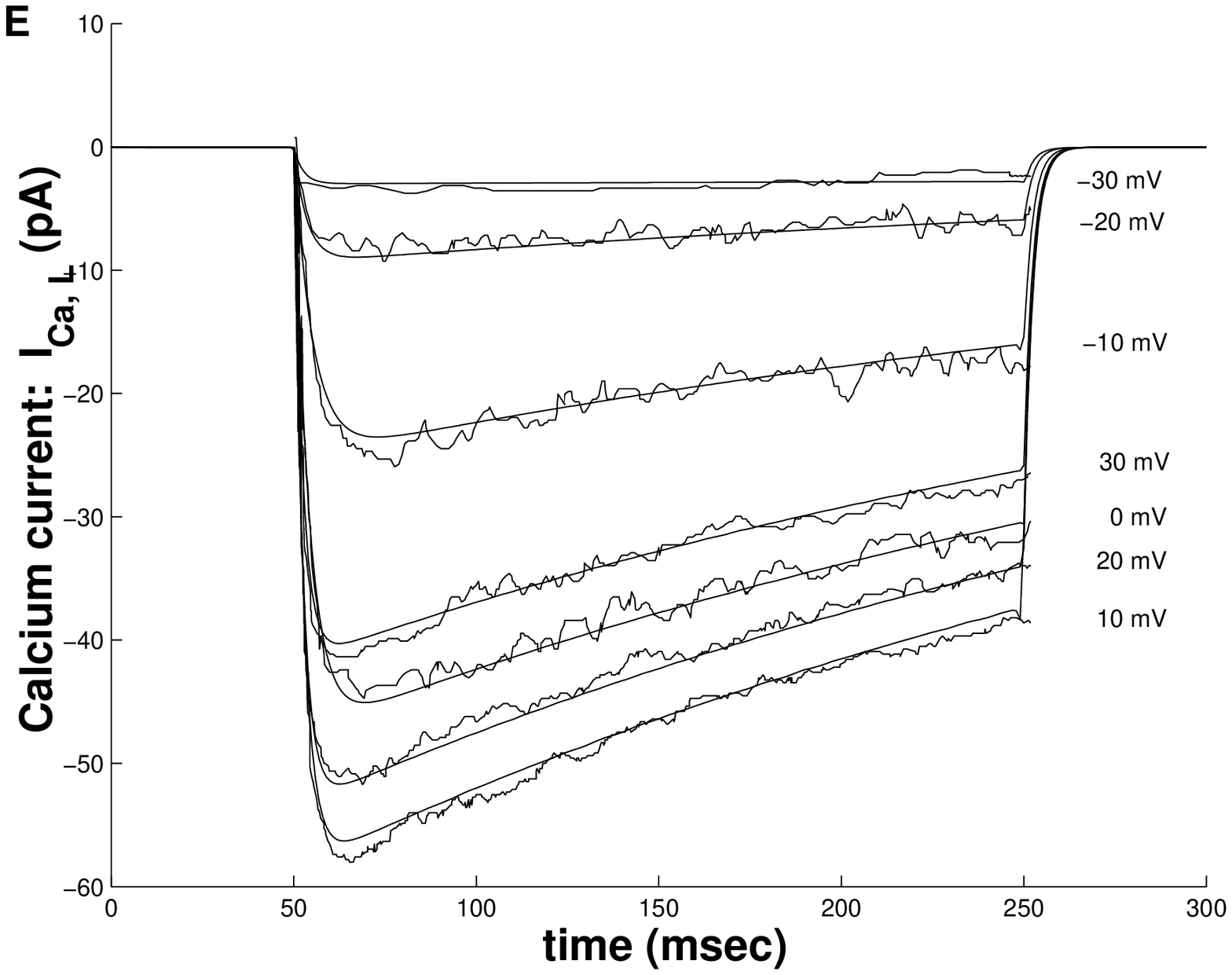}
  \caption{\label{figca} Characteristics of the L-type calcium channel. A. The steady-state activation ($\bar{d}_L$) and inactivation ($\bar{f}_L$) gating variables. B and C. Voltage-dependent time constants $\tau_d$ and $\tau_{f1}$ for activation and fast inactivation, respectively. D. Relationship between the peak $I_{\rm Ca,L}$ and the membrane potential. Data ($\ast$) from Rubart {\it et al.}~\cite{rub:96} and ($\circ$) from Langton~\cite{lan:93a}. E. Model-generated fits to measured voltage-clamp data from Rubart {\it et al.}~\cite{rub:96}. Voltage clamp (for a duration of 200 ms) steps are from -30 to 30 mV in 10 mV steps. The holding potential is -60 mV.}
\end{center}
\end{figure}

\subsubsection{Voltage-gated calcium current $I_{\rm Ca,L}$}
Both L-type and T-type \Ca \ currents have been described in whole-cell voltage-clamp studies on vascular smooth muscle~\cite{dav:99,htz:96,jrm:98,dav:92a,nel:90}. Regarding the time scales of \Ca \ channel activities, the L-type stands for a long-lasting inactivation, while the T-type stands for a transient activation and inactivation. However, unlike the L-type calcium current, the T-type has been found in some arterial smooth muscle cells, but not all~\cite{nel:90}. Single channel studies indicate that conductance of the L-type channel is approximately three times larger than that of the T-type channel (T-type, 8 pS; L-type, 25 pS~\cite{adh:95}). Our study is focused on cerebrovascular smooth muscle cells and for these cells, we assume that the T-type channel is weakly expressed or is absent for the following reasons: (1) Voltage clamp studies of isolated vascular smooth muscle cells indicate that activation and inactivation of T-type \Ca \ current $I_{\rm Ca,T}$ occurs over a voltage range that lies 20-40 mV negative to the threshold for the L-type channel current $I_{\rm Ca,T}$~\cite{adh:95}, which suggests that $I_{\rm Ca,T}$ may not be responsible for sustained smooth muscle contraction. (2) The amplitude of T-type current has been reported to be around 10-20\% of that of $I_{\rm Ca,L}$ and it is insensitive to dihydropyridine (DHP), which strongly inhibits maintained arterial tone~\cite{nel:90}. (3) The T-type channels are found in relatively high density in spontaneously active smooth muscle types and are associated with generating action potentials~\cite{htz:96}, but in contrast action potentials are not generated in cerebrovascular smooth muscle cells, suggesting that the role by the T-type channels were minimal.

The current $I_{\rm Ca,L}$ is assumed to be the only current underlying sustained \Ca \ entry and maintained contraction in cerebrovascular smooth muscle cells. Due to the strong voltage-dependence of $I_{\rm Ca,L}$, a membrane depolarization can modulate \Ca \ entry and consequently increase \Cai. At potentials in the range -60 to -40 mV, steady-state currents could be detected for minutes~\cite{nel:90,rub:96}, which suggests that $I_{\rm Ca,L}$ may provide sustained \Ca \ influx in the physiological range of membrane potentials above the resting state.

Figure~\ref{figca}A compares model-generated and measured steady-state voltage-dependent activation~\cite{lan:93a,rub:96} and inactivation characteristics for the $I_{\rm Ca,L}$. The model equations used for $I_{\rm Ca,L}$ are given in Appendix I: Table~\ref{tabica}, where the voltage- and time-dependent activation and inactivation variables are defined as $d_L$ and $f_L$, respectively. Figs.~\ref{figca}B and C show the voltage-dependence of the associated activation and inactivation time constants, respectively.  These time constants are modeled by Gaussian functions with an offset. We employ two time constants in describing the inactivation process for $I_{\rm Ca,L}$, since Rubart {\it et al.}~\cite{rub:96} report that both a fast and a slow component of inactivation may exist, with the slow time constant having a magnitude larger than 2 seconds or having a partial non-inactivation. Data obtained using 200 ms voltage clamp pulses~\cite{rub:96} does not show long term current inactivation. Therefore, we assume that the $I_{\rm Ca,L}$ current has a non-activation (Table~\ref{tabica}). Fig.~\ref{figca}D shows peak instantaneous $I_{\rm Ca,L}$ as a function of membrane voltage. This current is maximum at 10 mV and reverses its polarity at 60 mV.

The parameters associated with the voltage-dependent steady-state activation and inactivation functions and the associated time constants are identified by using a nonlinear least square optimization method (i.e., Levenberg-Marquardt algorithm)~\cite{nrc:93}. Fig.~\ref{figca}E compares model-generated and experimentally measured voltage-clamp dataset of Rubart {\it et al.}~\cite{rub:96} for $I_{\rm Ca,L}$.

\subsubsection{Stretch-sensitive current $I_{\rm M}$}
Stretch-sensitive ionic membrane currents have been shown to be present in arterial smooth muscle cells~\cite{dav:92b,dav:99,dav:92a}. When activated, these currents produce significant changes in membrane potential. In vascular smooth muscle cells, the stretch-sensitive membrane current $I_{\rm M}$ has a reversal potential about -15 mV~\cite{dav:92b}. We assume that the lumped channel conducting $I_{\rm M}$ has ionic components of ${\rm K}^+$, ${\rm Na}^+$ and \Ca \ (with selectivity order: ${\rm K}^+ > {\rm Na}^+ > {\rm Ca}^{2+}$),
and we express each of these components using a modified Goldman-Huxley-Katz equation:
\begin{eqnarray}
I_M & = & I_{\rm M,K}+I_{\rm M,Na}+I_{\rm M,Ca} \\
I_{\rm M,K} & = & \beta P_K\frac{[{\rm K}^+]_o-[{\rm K}^+]_ie^{V_mF/RT}}{1-e^{V_mF/RT}} \\
I_{\rm M,Na} & = & \beta P_{\rm Na}\frac{[{\rm Na}^+]_o-[{\rm Na}^+]_ie^{V_mF/RT}}{1-e^{V_mF/RT}} \\
I_{\rm M,Ca} & = & \beta z^2P_{\rm Ca}\frac{[{\rm Ca}^{2+}]_o-[{\rm Ca}^{2+}]_ie^{V_mzF/RT}}{1-e^{V_mzF/RT}} \\
P_m & = & \frac{1}{1+e^{-(\sigma-\sigma_{1/2})/k_{\sigma}}} \ ,
\end{eqnarray}
where the coefficient $\beta\equiv A_mP_mF^2V_m/(RT)$. The quantity $A_m$ is the surface area of the smooth muscle cell. We further assume that: (1) the geometry of the cerebrovascular smooth muscle cell is cylindrical; (2) the cell volume is 1.0 pl; and (3) the length of a cell is 150 ${\mu}$m. These geometric values can be used to calculate the cell-surface area $A_m$. The constants $P_K$, $P_{\rm Na}$ and $P_{\rm Ca}$ represent the permeability of the lumped channel to ${\rm K}^+$, ${\rm Na}^+$ and \Ca, respectively. The permeability ratios for ${\rm K}^+$, ${\rm Na}^+$ and \Ca (3.0:2.0:1.0) are adopted from Davis {\it et al.}~\cite{dav:92b}, and the open channel probability $P_m$ is characterized by a Boltzmann's relationship that is a function of the stress $\sigma$ experienced by smooth muscle membrane ($\sigma_{1/2}=204$ mmHg and $k_{\sigma}$ = 2.03 mmHg). The stress $\sigma$ is measured as force per unit cross-section area. Davis {\it et al.} reported that a 10-15\% stretch leads to a 40-50 pA whole-cell inward current through stretch-sensitive ion channels~\cite{dav:92b}.

\begin{figure}[t]
\begin{center}
  \includegraphics[scale=0.4]{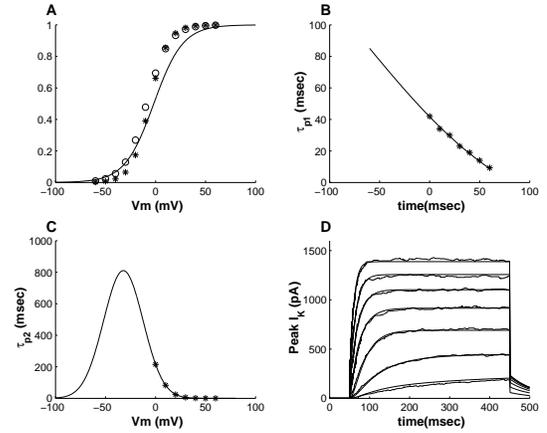}
  \caption{\label{fig4} Characteristics of voltage-dependent potassium channel (i.e., the delayed rectifier). A. Steady-state activation gating variable $\bar{p}_K$ compared to data from ($\ast$) Volk {\it et al.}~\cite{shi:91} and ($\circ$) Nelson {\it et al.}~\cite{nel:94} B. and C. Activation time constants $\tau_{\rm p1}$ and $\tau_{\rm p2}$, respectively, compared with data ($\ast$) from Volk {\it et al.}~\cite{shi:91} D. Model generated fits to measured voltage-clamp data (Volk {\it et al.}~\cite{shi:91}.) Voltage clamps (with a duration of 400 ms) are applied from the holding potential (-60 mv) to 10-60 mV in steps of 10 mv.}
\end{center}
\end{figure}

\subsubsection{\Ca-activated ${\rm K}^+$ current}
In vascular smooth muscle cells, the outward current is attributed largely to the \Ca-activated ${\rm K}^+$ current ($I_{\rm K,Ca}$)
\cite{dav:99,nel:95,wang:93}. This current is activated in response to a large influx of \Ca as the result of membrane depolarization. An increase in $I_{\rm K,Ca}$ brings about a hyperpolarization due to ${\rm K}^+$ efflux, which limits the voltage-dependent \Ca \ , and leads to smooth muscle relaxation. Patch clamp studies \cite{wang:93} have shown that single \Ca-activated ${\rm K}^+$ channels in smooth muscle cell isolated from rat cerebral arteries have a mean conductance of 207 pS. The time course of channel opening consists of two distinct components, {\it i.e.}, fast and slow \Ca-activated processes with mean time constants of 0.5 and  11.5 ms, respectively. The steady-state open probability of the channel is well-described by Boltzmann relationship that is shifted leftward by a 45mV per decade increase in \Cai (see Table \ref{tabkca} for its mathematical description).

\subsubsection{Inward rectifier current $I_{\rm K_i}$}
The inward rectifier current ($I_{\rm K_i}$) plays an important role in establishing the resting membrane potential of arterial smooth
muscle cell \cite{hirst:88,nel:95}. Control of channel activity by the extracellular potassium concentration $[{\rm K}^+]_o$ is also a
distinguishing feature of this current. Studies by Quayle {\it et al.}\cite{jmq:96} characterize the maximum slope conductance as a function of $[{\rm K}^+]_o$ according to:
\begin{equation}
g_{\rm max, K_i}  =  G_{\rm K_i}{([{\rm K}^+]_o)}^{n_{\rm K_i}} \ ,
\end{equation}
where $G_{\rm K_i}$ and $n_{\rm K_i}$ are constants with typical values of 0.145 and 0.5, respectively for pig coronary arterial smooth
muscle~\cite{jmq:96}. We have used a similar relationship in our model of rat cerebral arteries, and have adjusted the parameters
of this equation to better characterize measured data from rat cerebral arteries~\cite{hirst:88}. The parameter values used are
given in Table~\ref{tabpara1}. The chord conductance $g_K=I/(V- E_K$) has a Boltzmann relationship with membrane potential, and the mid-potential $V_{1/2,K_i}$ is a function of $[{\rm K}^+]_o$ . To characterize this relationship, we used the mathematical representation proposed by Quayle {\it et al.}~\cite{jmq:96}
\begin{equation}
V_{1/2,K_i}=A\log_{10} {[K^+]_i}  +  B \ .
\end{equation}
Typical values for $A$ and $B$ are 25.19 mV/$\log_{10}$(mM) and 112.29 mV, respectively. When the extracellular concentration $[{\rm K}^+]_o$ increases, the Boltzmann curve is shifted rightward toward depolarized potentials (see Appendix~\ref{appx1}: Table~\ref{tabki}).

\begin{figure}[t]
\begin{center}
\includegraphics[scale=0.38]{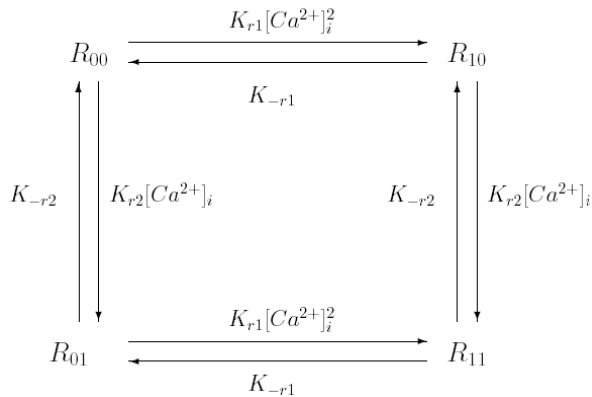}
\caption{\label{sr} A four-state kinetic model of ryanodine receptor activation controlling \Ca-induced \Ca \ release (CICR).The receptor has two regulatory sites that bind to \Ca \ for activation and inactivation.}
\end{center}
\end{figure}

\subsubsection{Delayed rectifier current $I_{\rm K}$}
The delayed rectifier current ($I_{\rm K}$) activates when the membrane is depolarized, and most vascular smooth muscle cells have this
current. It provides an outward current in response to membrane depolarization, which is important to counterbalance the effect of inward currents, thus limiting further membrane depolarization.

In our model, the steady-state activation process of $I_{\rm K}$ is voltage-dependent, and it is described by a Boltzmann function
(see Appendix I: Table~\ref{tabik}). Since the time constant associated with inactivation process is relatively large (2910 ms
at $V_m$=60 mV, Volk {\it et al.}~\cite{shi:91}), we do not include this slow process in our model. The activation process is characterized by two exponential components with different time constants. The voltage-dependent steady-state activation function is shown in Fig.~\ref{fig4}A, whereas the time constants of activation $\tau_{\rm p1}$ and $\tau_{\rm p2}$ are shown in Figs.~\ref{fig4}B and C, respectively. Model-generated fits to measured voltage clamp data (Volk {\it et al.}~\cite{shi:91}) are shown in Fig.~\ref{fig4}D. The half-activation voltage ($V_{1/2,K}$) and slope factor ($k$) of the Boltzman function, as well as, the  parameters associated with
voltage-dependent time constants are determined by using a non-linear least square parameter estimation method~\cite{nrc:93}. The values used in the model for $V_{1/2,K}$ (-1.77 mV) and $k$ (14.52 mV) are in good agreement with measurements by Volk {\it et al.}~\cite{shi:91} ($V_{1/2,K}=-6$ mV, $k=9$ mV) and by Nelson {\it et al.}~\cite{nel:94} ($V_{1/2,K}=-9$ mV, $k=11$ mV).

\subsubsection{${\rm Na}^+$/\Ca \ exchanger and membrane pumps}
Our model of the sarcolemma includes membrane currents associated with the ${\rm Na}^+/{\rm Ca}^{2+}$ exchanger ($I_{\rm NaCa}$), the ${\rm Na}^+/{\rm K}^+$ pump ($I_{\rm NaK}$ ), and the \Ca pump ($I_{\rm CaP}$). ${\rm Na}^+/{\rm Ca}^{2+}$ exchange is a counter-transport system that translocates ${\rm Na}^+$ and \Ca \ across the plasma membrane. This transporter has been well investigated and modeled for cardiac muscle cells~\cite{jwc:96}. Studies of this system in smooth muscle cells have also been reported \cite{meo:94}. The ${\rm Na}^+/{\rm Ca}^{2+}$ exchange system mediates either \Ca \ efflux in exchange for ${\rm Na}^+$ uptake or ${\rm Na}^+$ efflux in exchange for \Ca \ uptake depending on membrane potential ($V_m$) and the gradients for \Ca \ and ${\rm Na}^+$. The stoichiometry of
the exchange appears to be $3{\rm Na}^+:1{\rm Ca}^{2+}$~\cite{mcn:94}.

As an active transport system, the ${\rm Na}^+/{\rm K}^+$ pump is stimulated by the ${\rm Na}^+/{\rm K}^+$ ATPase and powered by hydrolysis of ATP to move ${\rm Na}^+$ and ${\rm K}^+$ ions against their electrochemical gradients. The stoichiometry of this active transport system is $3{\rm Na}^+:2{\rm K}^+$ and its ion transfer characteristic has a sigmoidal dependence on membrane potential~\cite{meo:94}. The ${\rm Na}^+/{\rm K}^+$ pump is electrogenic and provides a net outward current that helps to maintain the ${\rm Na}^+$ and ${\rm K}^+$ concentration gradient across the cell membrane.

Activity of the \Ca \ pump is regulated by calmodulin (CM), which serves to stimulate \Ca \ extrusion and \Ca-ATPase activity~\cite{meo:94}. In our study, we use a simple sigmoidal relationship to model the coupling between the whole cell \Ca \ pump
current and intracellular concentration of cytosolic \Ca, and assume that the CM gating kinetics of the \Ca \ pump are very fast
so that the binding process can be considered instantaneous (see Appendix: Table~\ref{tabpex}). Some studies report that the
$I_{\rm NaCa}$ does not play a significant role as a \Ca-removal mechanism in the smooth muscle cells of rat cerebral arteries~\cite{isen:91,mcn:98,nel:90}.

\subsubsection{CICR mechanism of sarcoplasmic reticulum}
In cerebrovascular smooth muscle cells, there is no general agreement regarding the importance of the sarcoplasmic reticulum (SR) in regulating \Cai. One group maintains that the \Ca \ transient is the result of sarcolemmal $I_{\rm Ca,L}$ influx, and suggest that local \Ca-release from ryanodine-sensitive \Ca \ into confined spaces regulates membrane potential via activation of $I_{\rm K,Ca}$~\cite{nel:00,nel:98b}. Others maintain that, in addition, there is a significant contribution to global intracellular \Cai \ by SR \Ca-release~\cite{mcn:97}. We model \Ca-induced \Ca-release (CICR) using a multiple-state kinetic model, which is similar to models developed for the SR of the cardiac ventricular cell~\cite{fab:92,stn:99,tang:94}. As shown in Fig.~\ref{sr}, the gating of a lumped ryanodine-sensitive \Ca \ release channel is regulated by the \Ca-dependent receptor kinetics. We assume that the secondary release of \Ca \ by SR affects the cytosolic \Ca \ concentration directly. The following differential equations describe this dynamic behavior of the different states of the ryanodine receptor (RyR):
\begin{widetext}
\begin{eqnarray}
\label{eqryan02}
\small
\begin{pmatrix}
      \frac{dR_{00}}{dt}\\
      \frac{dR_{10}}{dt}\\
      \frac{dR_{11}}{dt}\\
      \frac{dR_{01}}{dt}
\end{pmatrix} =
\begin{pmatrix}
   -(K_{\rm r1}[{\rm Ca}^{2+}]_i+K_{\rm r2})[{\rm Ca}^{2+}]_i & 0 & 0 & 0 \\
   K_{\rm r1}[{\rm Ca}^{2+}]_i^2 & -K_{\rm -r1}-K_{\rm r2}[{\rm Ca}^{2+}]_i & K_{\rm -r2} & 0 \\
   0 & K_{\rm r2}[{\rm Ca}^{2+}]_i & -K_{\rm -r1}-K_{\rm -r2} & K_{\rm r1}[{\rm Ca}^{2+}]_i^2 \\
   K_{\rm r2}[{\rm Ca}^{2+}]_i & 0 & K_{\rm -r1} & -K_{\rm -r2}-K_{\rm r1}[{\rm Ca}^{2+}]_i^2
\end{pmatrix}
\begin{pmatrix}
   R_{00}\\
   R_{10}\\
   R_{11}\\
   R_{01}
\end{pmatrix}
\end{eqnarray}
\end{widetext}
subject to the mass conservation constraint:
\begin{equation}\label{eqryan01}
R_{00}+R_{01}+R_{10}+R_{11}=1 \ ,
\end{equation}
where, the four different states of ryanodine receptor fractions are denoted as $R_{00}$: free receptors; $R_{10}$: receptors with \Ca \ bound to activation sites; $R_{01}$: receptors with \Ca \ bound to inactivation sites; and $R_{11}$: receptors with both activation and inactivation sites bound by \Ca. The open probability of the \Ca \ release channel controlled by this receptor mechanism is governed by the state variable $R_{10}$. The rates of \Ca \ binding to activation and inactivation sites are independent. Binding to activation sites is much faster than to inactivation sites (e.g., $K_{\rm r1}$ is much larger than $K_{\rm r2}$). In the model, there are two \Ca \ sites used for activation and one for inactivation and a second-order gating of release currents ($I_{\rm rel}$) by $R_{10}$ is applied. The above four ordinary differential equations (\ref{eqryan02}) finally reduce to a 3rd order system due to the algebraic constraint given in Eq.~(\ref{eqryan01}).

The SR itself is modeled as two-compartment (release/uptake) store with a fluid connection. A diffusion current $I_{tr}$ flows
between each the compartments, as shown in Fig.~\ref{fig1}. The SR is refilled via active pumping of \Ca \ into an uptake compartment, whereas the release compartment is responsible for storage of \Ca \ prior to release into the cytosolic space upon activation of the RyR. Intra-store \Ca \ balances and current fluxes are expressed as:
\begin{eqnarray}
I_{\rm up} & = & \bar{I}_{\rm up}\frac{[{\rm Ca}^{2+}]_i}{[{\rm Ca}^{2+}]_i + K_{\rm m,up}} \\
I_{\rm tr} & = & \frac{([{\rm Ca}^{2+}]_u- [{\rm Ca}^{2+}]_r)(2F\cdot
{\rm vol}_u)}{\tau_{\rm tr}} \\
I_{rel} & = & R_{10}^2([{\rm Ca}^{2+}]_{r}-[{\rm Ca}^{2+}]_i)\frac{(2F\cdot{\rm vol}_{r})}{\tau_{\rm rel}} \\
\frac{d[{\rm Ca}^{2+}]_{\rm up}}{dt} & = & \frac{I_{\rm up}-I_{\rm tr}}{2F\cdot{\rm vol}_u} \\
\frac{d[{\rm Ca}^{2+}]_{r}}{dt} & = & \frac{I_{\rm tr}-I_{\rm rel}}{2F\cdot{\rm vol}_r} \ ,
\end{eqnarray}
where the uptake current $I_{\rm up}$ is regulated by \Ca-ATPase activity, and the \Ca release current $I_{\rm rel}$ is gated by the activation state $R_{10}$ of ryanodine receptor model and the \Ca \ concentration gradient across the SR membrane. Description and numerical value of parameters associated with the model are listed in Table~\ref{tabsr}.

\subsubsection{Ionic balances of intracellular materials}
We have also developed a fluid compartment to account for changes in the cytosolic concentrations of ${\rm Na}^+$, ${\it K}^+$ and \Ca. This model includes a description of the geometry of smooth muscle cell, as well as \Ca \ buffering by calmodulin (CM). Since the
calcium-calmodulin complex (CaCM) plays such a vital role in regulating the myosin phosphorylation process, we distinguish this buffering process by modeling it separately. \Ca \ buffering by other elements such as the mitochondria, and the fluorescent indicator dye sensitive to \Cai, are lumped together as ``other" buffering processes ($[B_F]$, equation (\ref{eq:BF})). Material balance equations for intracellular ions are given in Appendix: Table \ref{tabmb}.  The extracellular ionic concentrations are assumed to be constant ($[{\rm Na}^+]_o = 140$ mM, $[{\rm K}^+]_o = 5$ mM and $[{\rm Ca}^{2+}]_o=2$ mM). Calmodulin has four free sites for binding \Ca. The average binding affinity of calmodulin sites for \Ca \ can be expressed as a dissociation constant ($K_d$) of 2.6$\times$10$^{-7}$ M based on the assumption that there are two high-affinity sites with $K_d=$2.0$\times$10$^{-7}$ M, and two low-affinity sites with $K_d=$2.0$\times$ 10$^{-6}$ M~\cite{kao2:97,wang:95}. The \Ca \ buffering by calmodulin and other \Ca \ binding ($[B_F]$) elements in the cytosol are described as first-order dynamic processes, which are represented by the following differential equation:
\begin{eqnarray}
\frac{d[S_{\rm CM}]}{dt} & = & k_{-1}[{\rm CaS}_{\rm CM}]-k_1[{\rm Ca}^{2+}]_i[S_{\rm CM}] \\
{[{\rm CaS}_{\rm CM}]} & = & [{\bar{S}}_{\rm CM}]-[S_{\rm CM}] \\
\frac{d[B_F]}{dt} & = & k_{-d}[{\rm CaB}_F]-k_d[{\rm Ca}^{2+}]_i[B_F] \label{eq:BF}\\
{[{\rm CaB}_F]} & = & [{\bar{B}}_F]-[B_F] \ ,
\end{eqnarray}
where $[S_{\rm CM}]$ represents the concentration of free calmodulin sites for \Ca binding and $[{\bar{S}}_{\rm CM}]$ represents the total concentration of calcium binding sites available for \Ca. The constants $k_{-1}$ and $k_1$ are kinetic rate constants for \Ca \ uptake and release, respectively, and the ratio of $k_{-1}$ to $k_1$ is equal to dissociation constant. Similar analysis is applied to the \Ca \ buffering process by other nonspecific proteins as described in Eq.~(\ref{eq:BF}).

\begin{figure}[t]
\begin{center}
\includegraphics[scale=0.4]{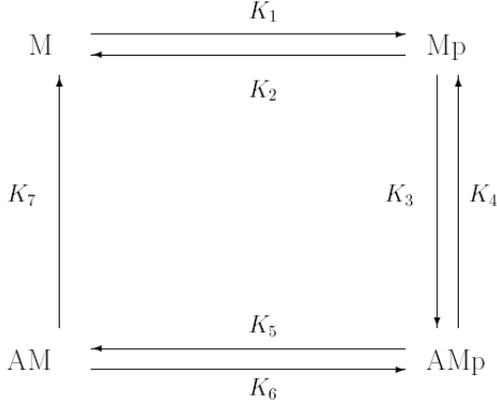}
\caption{\label{mult1} A four-state kinetic model of CaCM dependent myosin phosphorylation and cross--bridge formation. This figure is adopted from Hai and Murphy~\cite{hai:88a,hai:88b}.}
\end{center}
\end{figure}

\subsection{Kinetic model for smooth muscle cell contraction}
Interaction between actin and myosin filaments provides the molecular basis for smooth muscle contraction which is regulated by intracellular \Ca. Contractile mechanism of smooth muscle exhibits features which are similar to as well as distinguishing from striated muscle. A strong difference in smooth and striated muscle is that tension can be maintained in smooth muscle cells when \Cai and the level of phosphorylation are relatively low. This observation has led to the latch-bridge hypothesis of vascular smooth muscle contraction \cite{mur:90}. Although other mechanisms may co-exist for the regulation of contractile strength in smooth muscle ({\it e.g.}, protein kinase C may activate force generation with small or no changes in \Cai \cite{bar:96}), we assume that CaCM-MLCK dependent activation is the only mechanism present, and employ the four-state kinetic model developed by Hai and Murphy (H-M) \cite{hai:88a,hai:88b} to describe myosin phosphorylation and latch bridge formation.

Figure~\ref{mult1} is a depiction of H-M model, which consists of four fractional species: free cross-bridges (M), phosphorylated
cross-bridges (Mp), attached phosphorylated, cycling cross-bridges (AMp) and attached dephosphorylated non-cycling cross-bridges (as
latch bridges (AM)).  The latch state has been found to be unique in smooth muscle, which endows the smooth muscle cell with the ability to maintain force when the level of calcium-dependent myosin phosphorylation is relatively low. The kinetics of these four species can be described by the following differential equations:
\begin{widetext}
\begin{eqnarray}
\begin{pmatrix}
  d{\rm M}/dt\\
  d{\rm Mp}/dt\\
  d{\rm AMp}/dt\\
  d{\rm AM}/dt
\end{pmatrix} =
\begin{pmatrix}
  -K_1 & K_2 & 0 & K_7 \\
  K_1 & -K_2-K_3 & K_4 &0 \\
  0 & K_3 & -K_4-K_5 & K_6 \\
  0 & 0 & K_5 & -K_6-K_7
\end{pmatrix}
\begin{pmatrix}
  {\rm M} \\
  {\rm Mp} \\
  {\rm AMp} \\
  {\rm AM}
\end{pmatrix}
\end{eqnarray}
\end{widetext}
subject to the constraint
\begin{equation}
 {\rm M+Mp+AMp+AM}=1 \ ,
\end{equation}
where the fraction of phosphorylated myosin is defined as the sum Mp+AMp, whereas attached cross--bridges are represented by the sum AMp+AM. In our modification of H-M model, the CaCM-dependence of myosin phosphorylation, rather than direct dependence on \Ca, is represented through rate constants $K_1$ and $K_6$ with an analytic relationship as described in Eq.~(\ref{eq:k16}). $K_1 = K_6$ is assumed, which means that the phosphorylation process has same rate in changing from M to Mp, as from AM to AMp. Thus, 
\begin{equation}\label{eq:k16}
K_1 = K_6 = \frac{[{\rm CaCM}]^2}{[{\rm CaCM}]^2+K_{\rm CaCM}^2} \ ,
\end{equation}
where $K_{\rm CaCM}$ is the half-activation constant for the CaCM-dependent phosphorylation rate constant. This sigmoidal relationship indicates the saturation of the rate constants $K_1$ and $K_6$ when the concentration of CaCM is large. We also assume the dephosphorylation process from Mp to M occurs at the same rate as from AMp to AM, {\it i.e.}, $K_2 = K_5$.

\subsection{Mechanical properties of the cell}

Experimental studies conducted by Warshaw and Fay on the length-force relationship of the single smooth muscle cell report that cross-bridge stiffness is between 1.2 and 1.5 times greater than the measured cell stiffness. They report that the elastic response originating within the muscle cell is attributed to cross-bridges acting in series with an elastic element that has an exponential length-force relationship~\cite{fay:83,fay:88b}. A viscous contribution to cell stiffness has also been detected in experiments where the phase shift between an applied cell length change and the observed change in force has been measured~\cite{fay:83,fay:88b}.

\begin{figure*}[htb]
\begin{center}
\includegraphics[scale=0.5]{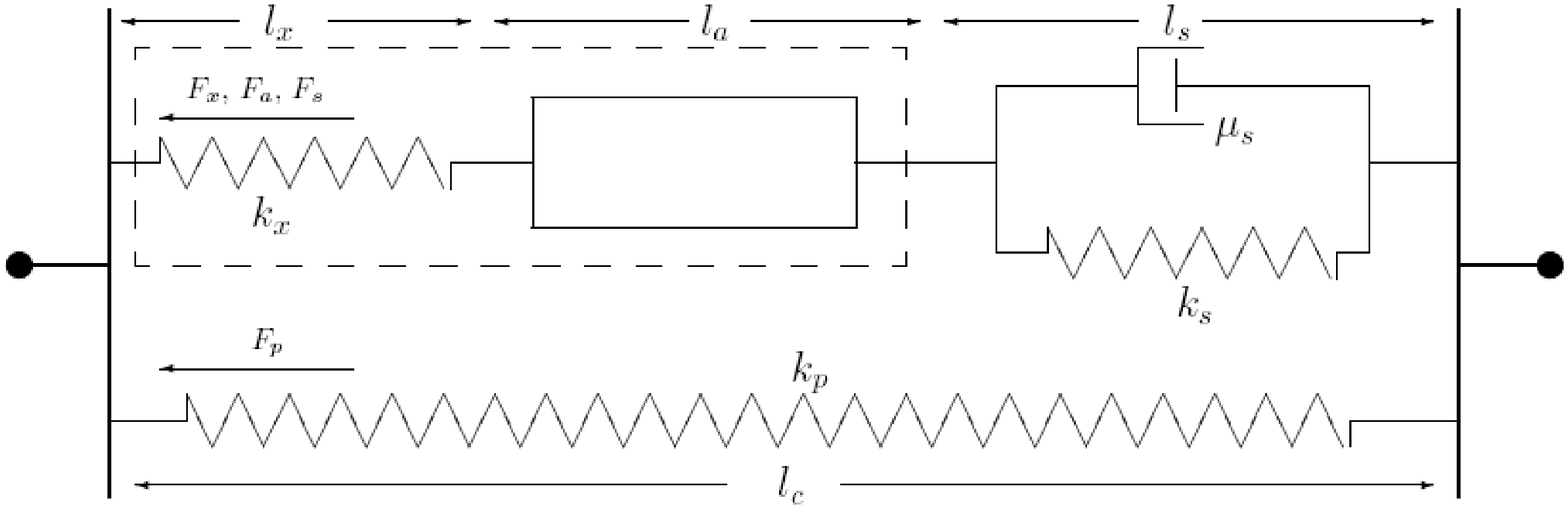}
\caption{\label{fmech} Mechanical model for the smooth muscle cell. The passive length-force properties of the cell are described by the spring with stiffness $k_p$, whereas the springs characterized by stiffness $k_x$ and $k_s$ denote the elasticity of attached cross bridge and series elastic element, respectively. The wall attachment coupling also has a viscous element characterized by the damping coefficient $\mu_s$. The active contractile element (length $l_a$) models cross-bridge actin filament interaction by the forces generated by cross-bridge cycling and sliding.}
\end{center}
\end{figure*}

A modified Hill model~\cite{ycf:93} is used to describe the mechanics of the cell (Fig.~\ref{fmech}). This model reflects the coupling between active the force generation by the cross-bridge mechanism and the mechanical coupling properties of the myofilaments and their viscoelastic attachment to structures in communication with the cell wall.

\subsubsection{Passive Force}
The passive elasticity of the cell ($k_p$) is modeled by an exponential force-length relationship used to represent the nonlinearity of the passive elasticity. The passive force is given by:
\begin{equation}
F_p = k_p\left[e^{\alpha_p\cdot(l_c/l_0-1)}-1\right] \ ,
\end{equation}
where $k_p$ is a constant and $l_0$ is the length at which the cell generates zero passive tension.

\subsubsection{Cross-bridge elasticity}
Due to the elasticity of attached cross bridges, cell tension can also be sustained by cross-bridges, and this length-dependent force is characterized by:
\begin{equation}\label{crosselas}
F_x = (k_{\rm x_1} \cdot {\rm AMp}+k_{\rm x_2} \cdot {\rm AM}) \cdot l_x \cdot
e^{-\beta{\left(l_a/l_{\rm opt}-1\right)}^2} \ ,
\end{equation}
where $l_x$ is the extension of cross--bridge and $l_a$ is the length of active element which indicates the overlap between the myosin and actin filaments. The optimal length ($l_{\rm opt}$) of the active element is the length at which maximum overlap occurs between the myosin and actin filaments, and it is also the length at which the cell develops maximum tension for the active state achieved. We approximate the effect of filament overlap in the development of active tension by the Gaussian function in Eq.(\ref{crosselas}). This equation for $F_x$ also reflects that both types of cross-bridges (cycling and latch bridges) contribute to the elasticity of the attached cross-bridges. The spring constants $k_{\rm x1}$ and $k_{\rm x2}$ denote the maximal stiffness that can be achieved by phosphorylated attached cross-bridges and latched cross-bridges, respectively. States AMp and AM are the outputs of the kinetic contractile model described in previous section, and indicate the distribution of attached cross-bridges and the state of interaction between myosin and actin filaments. Consequently, the effective stiffness contributed by phosphorylated cross-bridge is the product $k_{\rm x1}$ AMp, whereas that contributed by latch bridges is $k_{\rm x2}$ AM. Eq. (\ref{crosselas}) also indicates that the two types of attached cross-bridges with different effective stiffness contribute in parallel to the  total active force $F_x$ and the total stiffness is further modulated by cell contractile kinetics which result in a change in active element length $l_a$.

\subsubsection{Active force generation}
The active force generated by attached cross-bridge is given by:
\begin{equation}\label{crossfric}
F_a = \left[f_{\rm AMp}\cdot{\rm AMp}(v_{x}+\frac{dl_a}{dt}) + f_{\rm AM}\cdot
{\rm AM}\frac{dl_a}{dt}\right]e^{-\beta{\left(l_a/l_{\rm opt}-1\right)}^2} \ ,
\end{equation}
where $f_{\rm AMp}$ and $f_{\rm AM}$ are friction constants for phosphorylated cross bridges and latch bridges, respectively. In this formulation, we assume: (1) Only attached phosphorylated cross bridges exhibit cycling behavior which produces translational motion of the actin filament at a velocity $v_x$, and this is the only process contributing to active force generation. Latch bridges lose cycling capability, and their continuing attachment to the actin filament increases the stiffness of the cross-bridge coupling [Eq.~(\ref{crosselas})]; (2) Both species of attached cross bridges, phosphorylated and latch bridges, slide along the actin filament with velocity $dl_a/dt$ in response to cell stretch, release and contraction.

The first assumption indicates that populations of attached cross-bridges are heterogeneous, and that latch bridges act as an internal load on fast cycling cross-bridges~\cite{mur:83,dmw:90}. Note that both the total effective stiffness and the friction offered by the cross--bridges are subject to modulation by $l_a$, the length of the active element, through a Gaussian relationship that is expressed in Eqs. (\ref{crosselas}) and (\ref{crossfric}), respectively. The Gaussian function expresses the effect of the degree of overlap of the actin and myosin filaments.

\subsubsection{Series viscoelastic force}
The series coupling element is modeled as an elastic Kelvin-Voigt body that consists of a spring (by an elastic constant $k_s$) connected in parallel with a dashpot (by a frictional constant $\mu_s$). The total force across the body is given as the sum of elastic and viscous forces, where the spring is characterized by an exponential length-force relationship, and the force sustained by the series viscoelastic element is expressed as:
\begin{equation}
F_s = \mu_s\frac{dl_s}{dt}+k_s\cdot(e^{\alpha_s\cdot(l_s/l_{s0}-1)}-1) \ ,
\end{equation}
where the viscous coefficient $\mu_s$ and $k_s$ are constants, and $l_{s0}$ is the length at which the series elastic component sustains zero force.

At each instant of time, the force sustained by (a) the cross bridges, (b) the active force generation element, and (c) the series element is the same, {\it i.e.},
\begin{equation}
F_s=F_a=F_x \ .
\end{equation}
The total force sustained by the cell is:
\begin{equation}
F_t=F_a + F_p \ .
\end{equation}
The numerical values of parameters associated with this cell mechanics model are listed in Table~\ref{tabmech}.

Equations and parameters of the electrochemical model are listed in Table~\ref{tabpara1} through Table~\ref{tabpara2}. Initial conditions for typical model simulations are listed in Table~\ref{tabinit}. The complete model consists of 23 state variables.  Model parameters are identified by using numerical optimization methods with non-linear least square techniques~\cite{nrc:93} to fit experimental data. The model is implemented and the results are visualized by using the MATLAB programming language on an Intel-based PC platform.

\section{Results}
We test the cell model by subjecting it to conditions that emulate various experimental protocols used in testing the functional behavior of single smooth muscle cells, and compare model predictions with corresponding measured data from the literature. Agreement between model predictions and measured data over a wide range of testing protocols indicates the capability of this model to represent key mechanisms involved in excitation-contraction coupling, as well as, the general electromechanical behavior of this cerebrovascular smooth muscle cell.

\begin{figure}[htb]
\begin{center}
\includegraphics[scale=0.43]{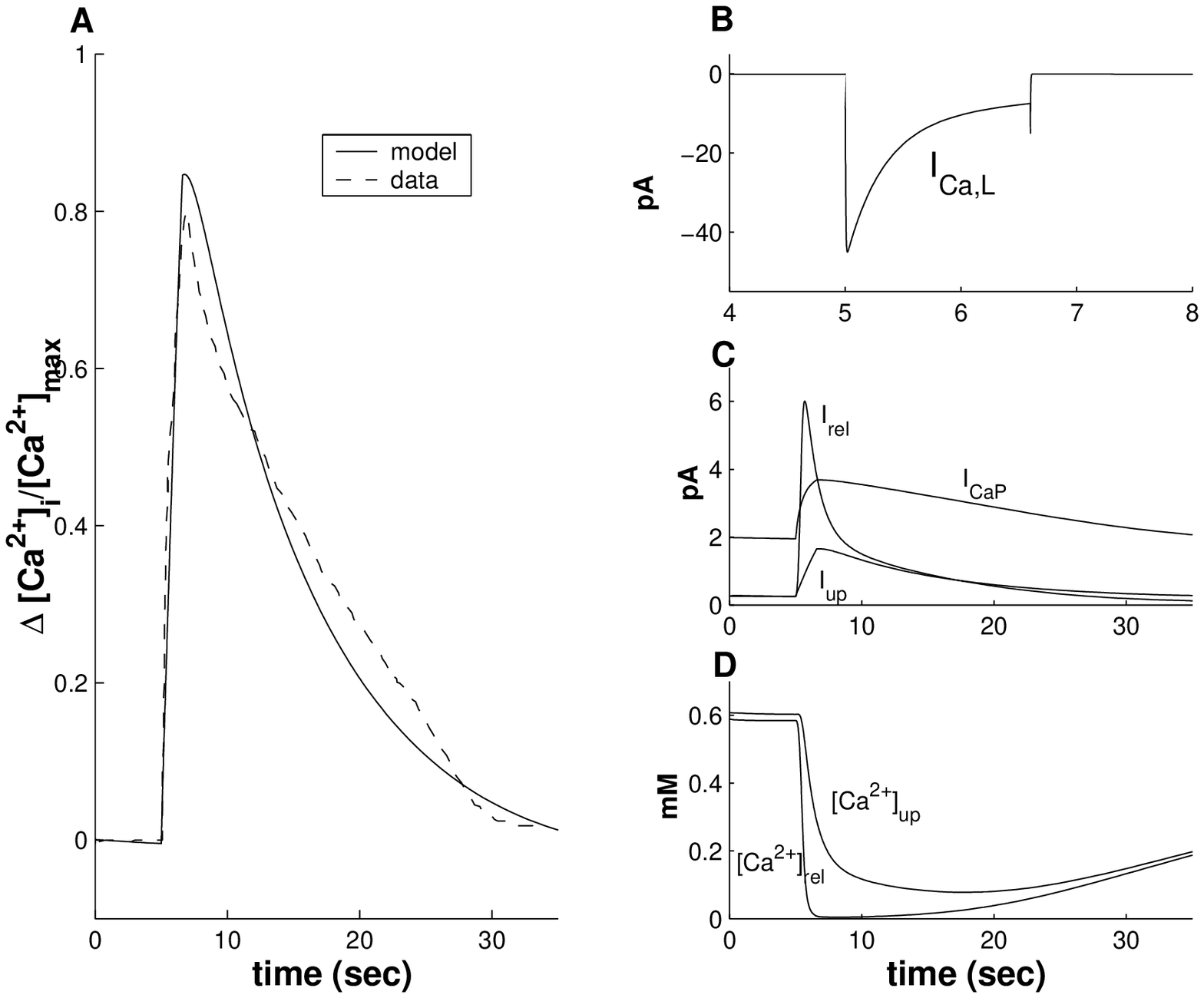}
\includegraphics[scale=0.43]{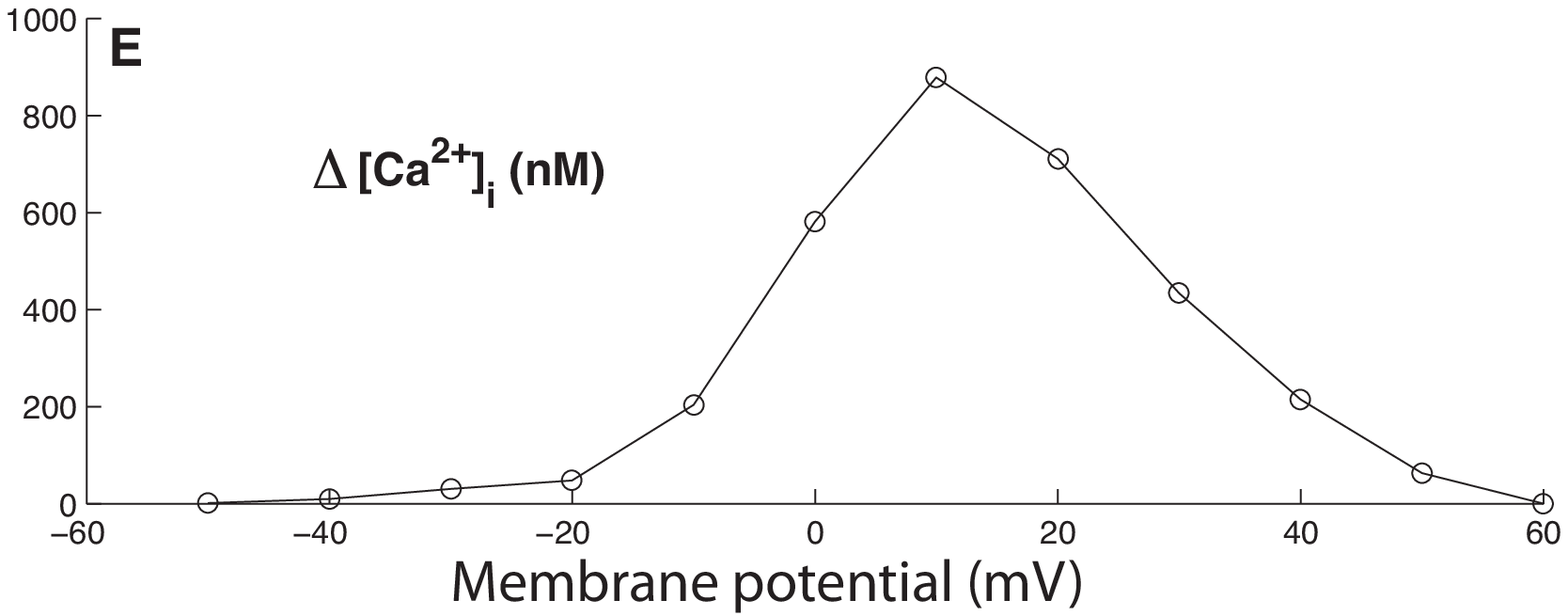}
\caption{\label{catrans} Calcium transient in response to a voltage pulse. A. Intracellular \Ca \ transient in response to 1.6 s step depolarization to 0 mV from a holding potential of -60 mV. Model-generated data is plotted with measured data from Kamishima {\it et al.}~\cite{mcn:97,mcn:98}). Here the change in \Cai for model-generated and measured data are each normalized by the maximum peak value
$[{\rm Ca}^{2+}]_{\rm max}$ encountered in the series of voltage clamp tests (see Fig.\ref{catrans}), which happens at 10 mV as shown in panel E. B. Voltage-dependent L-type calcium current $I_{\rm Ca,L}$ in response to voltage step (note, this panel has a different time
scale than others). C. Calcium removal currents: the sarcolemma \Ca-ATPase pump current $I_{\rm Ca,P}$ \ and the SR membrane \Ca-ATPase uptake current $I_{\rm up}$. D. \Ca \ concentrations in the uptake and release compartments of the SR. E. Voltage dependence of change in \Cai ($\Delta$\Cai) measured relative to baseline \Cai.}
\end{center}
\end{figure}

\subsection{Testing the electrochemical model}
\subsubsection{Calcium transient}
The \Ca \ transient is of importance for proper smooth muscle function, since it represents the input to the contractile mechanism.  In cerebrovascular smooth muscle, \Ca \ entry into the cytosol is mainly via the inward voltage-dependent membrane current $I_{\rm Ca,L}$ \cite{nel:90}. The principal \Ca-removal mechanisms at work in this cell type are the separate and distinct sarcolemmal and SR \Ca-ATPase pumps.  Kamishima {\it et al.}~\cite{mcn:98} report that in the rat cerebral arterial smooth muscle cell, the rate of \Ca \ removal exhibited by cells bathed in a ${\rm Na}^+$ free solution was not significantly different from that observed in control cells. Studies by Ganitkevich {\it et al.}\cite{isen:91} on smooth muscle cells of guinea pig urinary bladder, also indicate that \Ca \ extrusion via $I_{\rm NaCa}$ is negligible in this cell type. Although our model contains a mathematical expression for $I_{\rm NaCa}$, the magnitude of the sarcolemmal pump current $I_{\rm Ca,P}$ and the SR uptake current $I_{\rm up}$ are made relatively large compared with $I_{\rm NaCa}$, and hence $I_{\rm NaCa}$ is considered small.

Calcium transients can be induced in several ways in smooth muscle cell. Here we focus on the voltage clamp method and emulate the experimental protocols used by Ganitkevich {\it et al.}~\cite{isen:91} and Kamishima {\it et al.}~\cite{mcn:97,mcn:98} to produce \Ca \ transients. Fig.~\ref{catrans}A compares the model-generated \Ca \ transient in response to a $1.6$ s voltage pulse from holding potential of -60 mV to a depolarized level of 0 mV. The measured data shown in this figure is from Kamishima {\it et al.}~\cite{mcn:97,mcn:98}. Figure~\ref{catrans}A shows that the depolarization pulse triggers an increase in cytosolic calcium
concentration (\Cai), whereas Figs.~\ref{catrans}B and C show that the leading edge of this \Ca \ transient is due mainly to
$I_{\rm Ca,L}$, but also to the SR release current $I_{\rm rel}$. Decay in the \Ca transient is mainly caused by the rapid decline in $I_{\rm Ca,L}$ coupled with the increase in sarcolemmal and SR uptake pump currents ($I_{\rm CaP}$ and $I_{\rm up}$, respectively). Fig.~\ref{catrans}D shows that $[{\rm Ca}^{2+}]$ within the SR compartments is quickly depleted with the onset of the voltage pulse and the SR \Ca \ release ($I_{\rm rel}$). The model predicts that the SR compartments are slow to refill after the depletion of the store (Fig.\ref{catrans}D).

Experiments by Ganitkevich {\it et al.}~\cite{isen:91} on single smooth muscle cells of the guinea-pig urinary bladder show that the voltage-dependence of peak \Ca \ transients resembles the bell-shaped peak I-V relationship of $I_{\rm Ca,L}$ (Fig.~\ref{figca}D). Fig.~\ref{catrans}E shows that the model-generated peak \Ca \ transients exhibit a bell-shape relationship with membrane potential, similar to that exhibited by $I_{\rm Ca,L}$. Both curves have maxima at 10 mV. The protocol followed by Ganitkevich {\it et al.} \cite{isen:91} to study the voltage-dependence of the \Ca \ transient, involves the delivery of a number of $1.6$ s depolarization pulses of different amplitudes over the range -40 mV $<V<$ +60 mV in 10 mV steps, starting at the holding potential of -60 mV. Cytosolic $I_{\rm Ca,L}$ and the \Cai transient are measured at each voltage step.

\begin{figure}[t]
\begin{center}
  \includegraphics[scale=0.4]{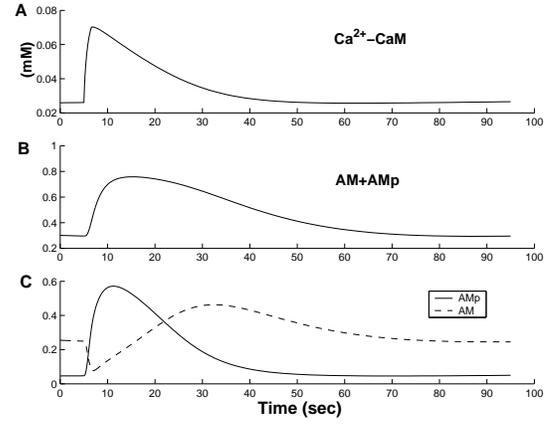}
  \caption{CaCM-dependent myosin phosphorylation and cross-bridge distribution. Calcium transient is elicited with 1.6 s voltage depolarization from -60 to 0 mV. A. Waveform of the concentration of CM sites occupied by \Ca. B. Total fraction of attached cross-bridge population (AM+AMp). C. Temporal distribution of attached cross-bridges in phosphorylated (AMp : solid line) and dephosphorylated states (AM: dashed line).}\label{fig:phos}
\end{center}
\end{figure}

\subsection{Testing the chemomechanical model}
\subsubsection{CaCM-dependent myosin phosphorylation}
Increases of intracellular \Ca \ initiate the smooth muscle contraction through the postulated ${\rm Ca}^{2+}->$CaCM$->$MLCK pathway
of myosin phosphorylation and cross-bridge formation. Model simulation of this cellular event is shown in Fig.~\ref{fig:phos}. A 1.6s membrane depolarization from -60 to 0 mV is applied to generate the cytosolic \Ca \ transient. Fig.~\ref{fig:phos}A shows the dynamics
of the concentration of the CM sites occupied by \Ca \ ion in response to the \Cai transient. Kinetic model simulation of CaCM activated myosin phosphorylation give the response of total attached cross-bridges population as shown in Fig.~\ref{fig:phos}B, whereas the distribution and transition of cross-bridges between phosphorylated state and latch state is shown in Fig.~\ref{fig:phos}C. The majority of attached cross-bridges at the resting level of -60mV are latch bridges. Phosphorylation of free myosin M and the transition AM$\rightarrow$AMp following the \Ca \ transient contributes to the rapid increase of the active cycling cross-bridge population. As the CaCM level decreases, the dephosphorylation process becomes dominant and the attached cross-bridge population tends to transfer back into the latch state.

\begin{figure}[t]
\begin{center}
\includegraphics[scale=0.4]{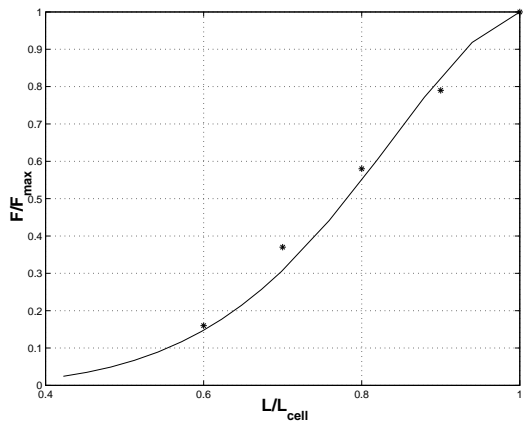}
\caption{\label{fig:lf} Length-force relationship in the isolated smooth muscle cell. Model-generated data compared with measured data (*) is from Harris {\it et al.}~\cite{dmw:91}.}
\end{center}
\end{figure}

\subsubsection{Length-force relationship}
Classic isometric test protocols have been applied to study the static length-force (L:F) relationship in isolated smooth muscle
cells~\cite{dmw:91}. Specifically, our model simulations employ a very slow ramp of length increase from $l_0$ (see Appendix~\ref{appx1}: Table~\ref{tabmech}) and the change in force is observed. In Fig.~\ref{fig:lf}, the L:F relationships predicted by the model are compared with measured data (Harris {\it et al.}~\cite{dmw:91} on toad stomach smooth muscle cells). For cell lengths below the optimal length (maximum overlap between myosin and actin filaments; $L_{cell}$ in Fig.~\ref{fig:lf}), there is an approximately linear relationship between length and force. Figure~\ref{fig:lf} shows that force tends to approach zero when cell length becomes less than 0.4 of $L_{cell}$, which agrees well with the measurements of Harris {\it et al.}~\cite{dmw:91}.

\subsubsection{Force-velocity relationship}
A classic isotonic quick-release technique is frequently applied to the study of the relationship between force and shortening
velocity in muscle. Measurements by Warshaw~\cite{dmw:87} on isolated single smooth muscle cells of toad stomach muscularis
shows that force-velocity relationship of the single cell agrees with the well-known hyperbolic description of Hill equation~\cite{hill:38}:
\begin{equation}\label{eq:hill}
(\frac{F}{F_{\rm max}} +
\frac{a}{F_{\rm max}})(V+b)=(1+\frac{a}{F_{\rm max}})b \ ,
\end{equation}
where $F_{\rm max}$ is maximum isometric active force, and $a$ and $b$ are constants.

\begin{figure}[t]
\begin{center}
\includegraphics[scale=0.36]{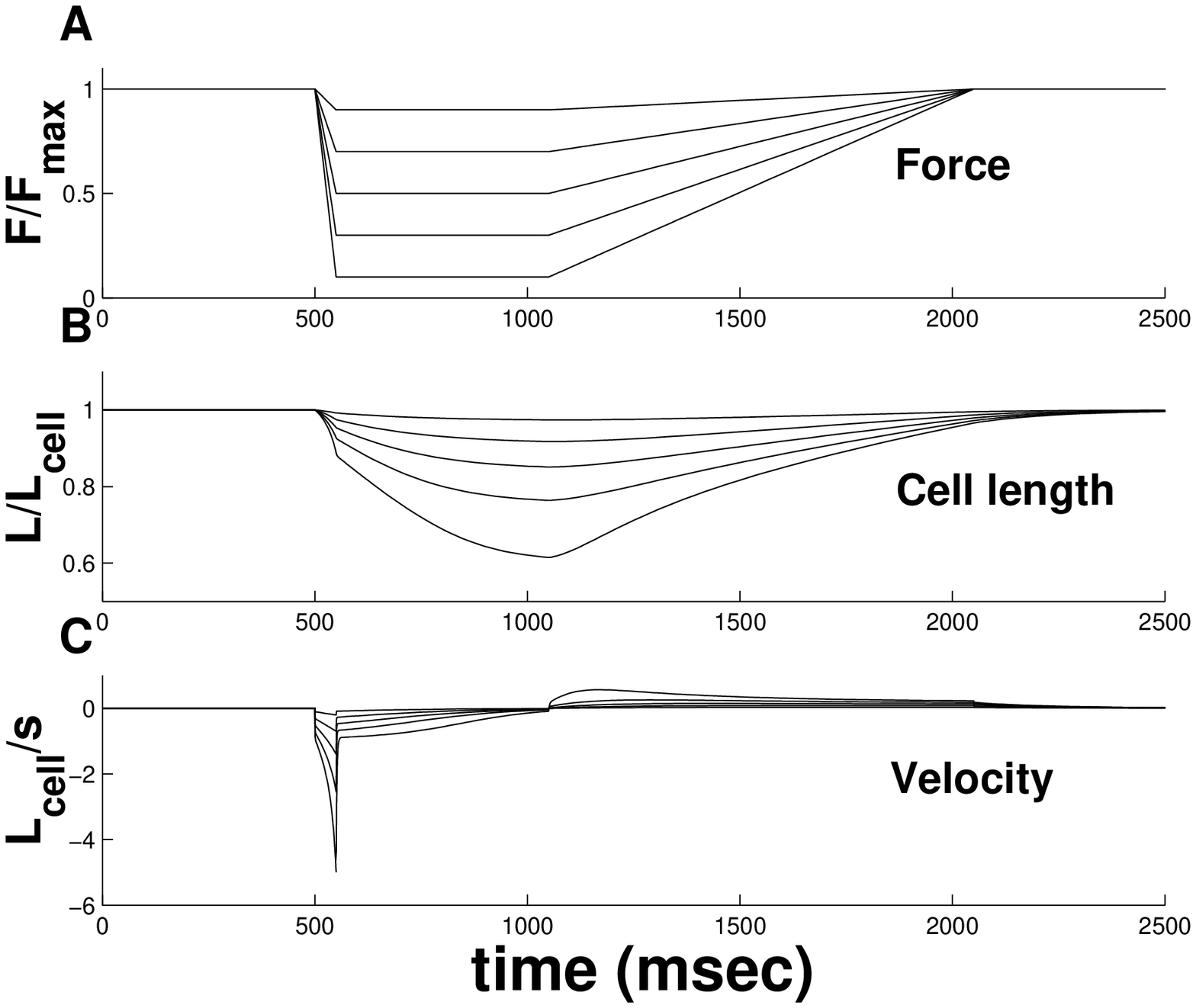}
\includegraphics[scale=0.36]{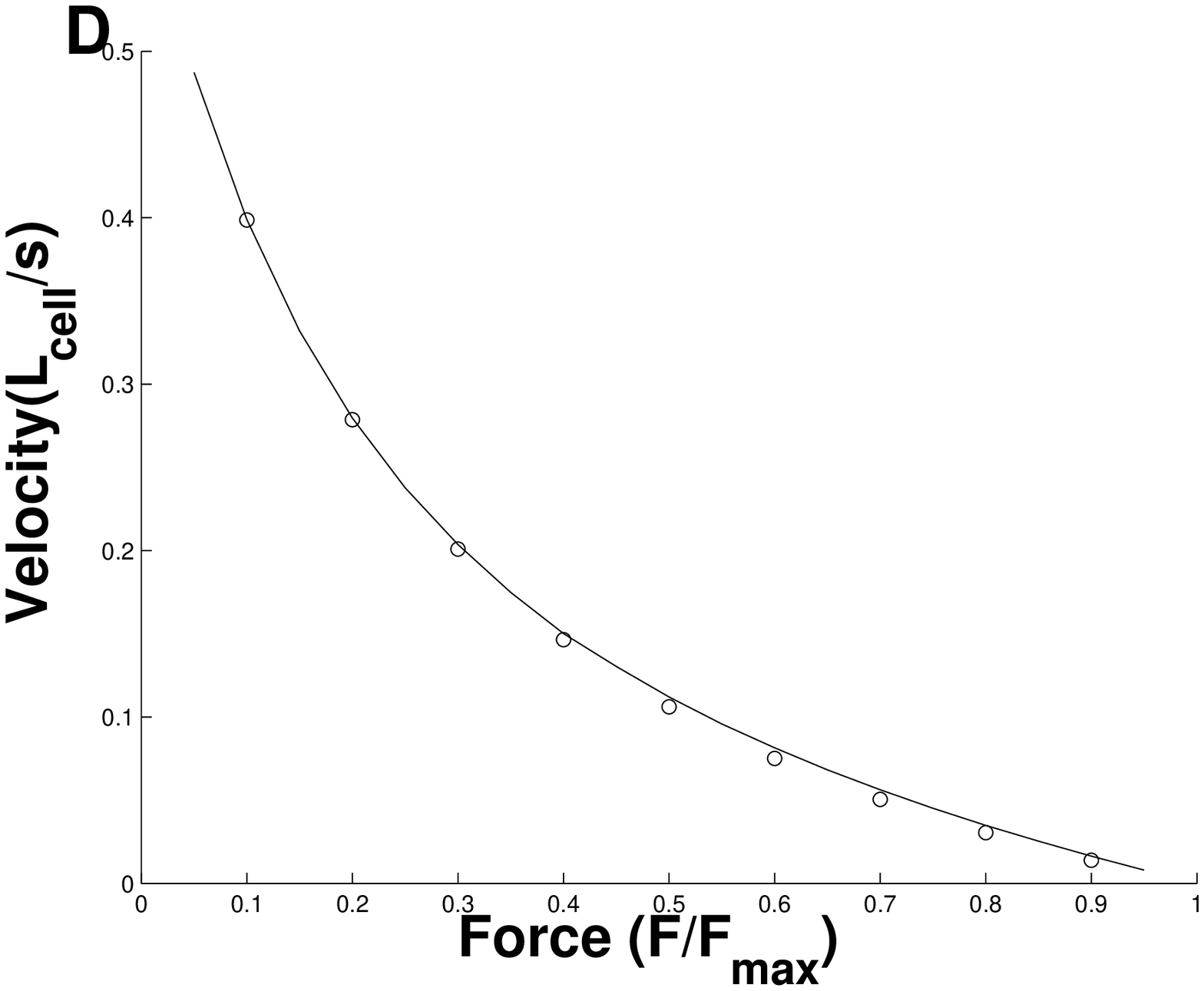}
\caption{\label{fig:fv} Length and velocity responses in isotonic tests of the model and force-velocity relationship.  A. Isotonic force release protocol: release magnitudes from 0.1 to 0.9$F_{\rm max}$ with step-size 0.2$F_{\rm max}$. Each force level is maintained for
500 ms. B. and C.  Length and velocity changes in response to the force changes, respectively. Note that shortening velocity in F:V
relationship should be the velocity taken after the isotonic release. $F_{\rm max}=2.13$ $\mu$N, and $l_{\rm cell}=123$ $\mu$m. D. Solid line: model simulation of F:V relationship. Data (o) from the Hill hyperbola [see Eq.~(\ref{eq:hill})] which was used by Warshaw to fit data from on toad stomach muscularis~\cite{dmw:87}, with $a/F_{\rm max}=$0.268 and $b = 0.163$ $L_{\rm cell}/s$.} 
\end{center}
\end{figure}

In isotonic tests, the cell is electrically activated~\cite{dmw:87}. The output of the contractile kinetic model is the sum of the attached cross-bridges (AM + AMp). Compared to the duration of AM+AMp in response to a calcium transient (typical duration: 50-70 s), the duration of the a isotonic test is much smaller (1.5-2 s). Thus, myosin phosphorylation kinetics are assumed to be relatively constant during isotonic conditions. Consequently, AM and AMp are set to constants during isotonic test procedures (typical values AM=0.3 and AMp=0.5; Fig.~\ref{fig:phos}). Model-generated length and velocity responses to quick releases in force are shown in Fig.~\ref{fig:fv}. The maximum isometric force $F_{max}$ is the force sustained by the cell before the initial force release, and it is a function of the
contractile kinetics and cell length($L_{\rm cell}$). $F_{\rm max}$ is used as the reference to a series of force releases, which have
various magnitudes ranging from 0.1 to 0.9 $F_{\rm max}$ with a step-size 0.2 $F_{\rm max}$. Force release to the test level is
complete within 50 ms for each trial. The test force level is maintained for 500 ms before the cell is brought back to $F_{\rm max}$ over a 1.0 second period (Fig.~\ref{fig:fv}A). Cell length and velocity responses are shown Figs.~\ref{fig:fv}B and C, respectively.

Figure~\ref{fig:fv}D shows model-predicted and measured data by Warshaw~\cite{dmw:87} for the force-velocity (F:V) relationship of the single smooth muscle cell. Warshaw reports that the F:V relationship can be fitted by Hill hyperbola as described in Eq.~(\ref{eq:hill}) with $a/F_{\rm max}=$0.268 and $b = 0.163L_{\rm cell}/s$~\cite{dmw:87}. Panel D also shows that our model can produce reasonable fits to measured F:V data, in that it provides good fit to the Hill equation, which was used by Warshaw~\cite{dmw:87} to fit their data.

\subsection{Complete model testing}
The full or complete model is now subjected to tests that highlight the relationships between membrane potential, cell strain, calcium transients force generation and mechanical response. Two stimulation protocols are applied: (1) membrane depolarization induced by voltage pulse and (2) strain change induced by cell stretch.

\subsubsection{Electrical forcing}
\begin{figure}[b]
\begin{center}
\includegraphics[scale=0.38]{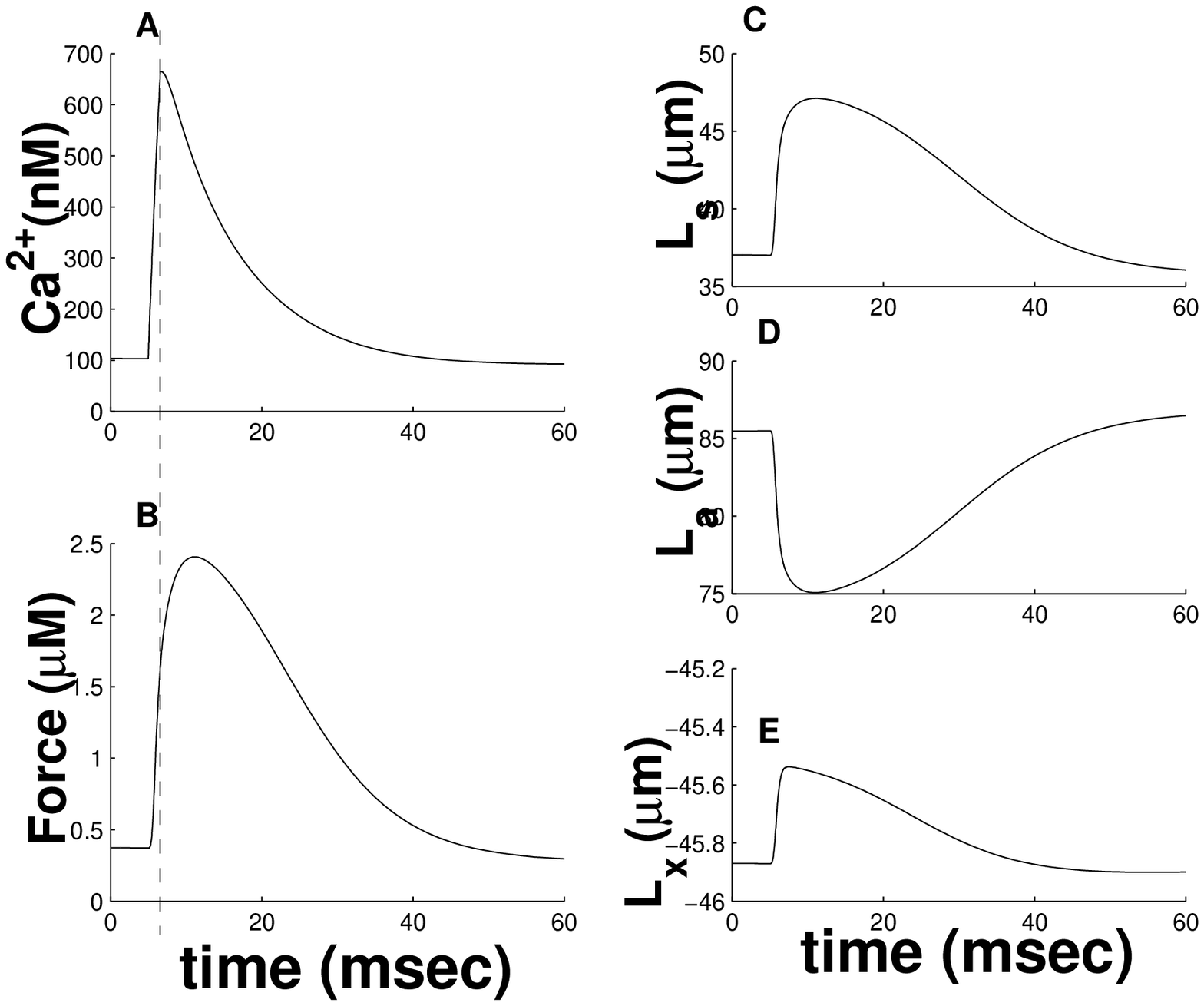}
\includegraphics[scale=0.38]{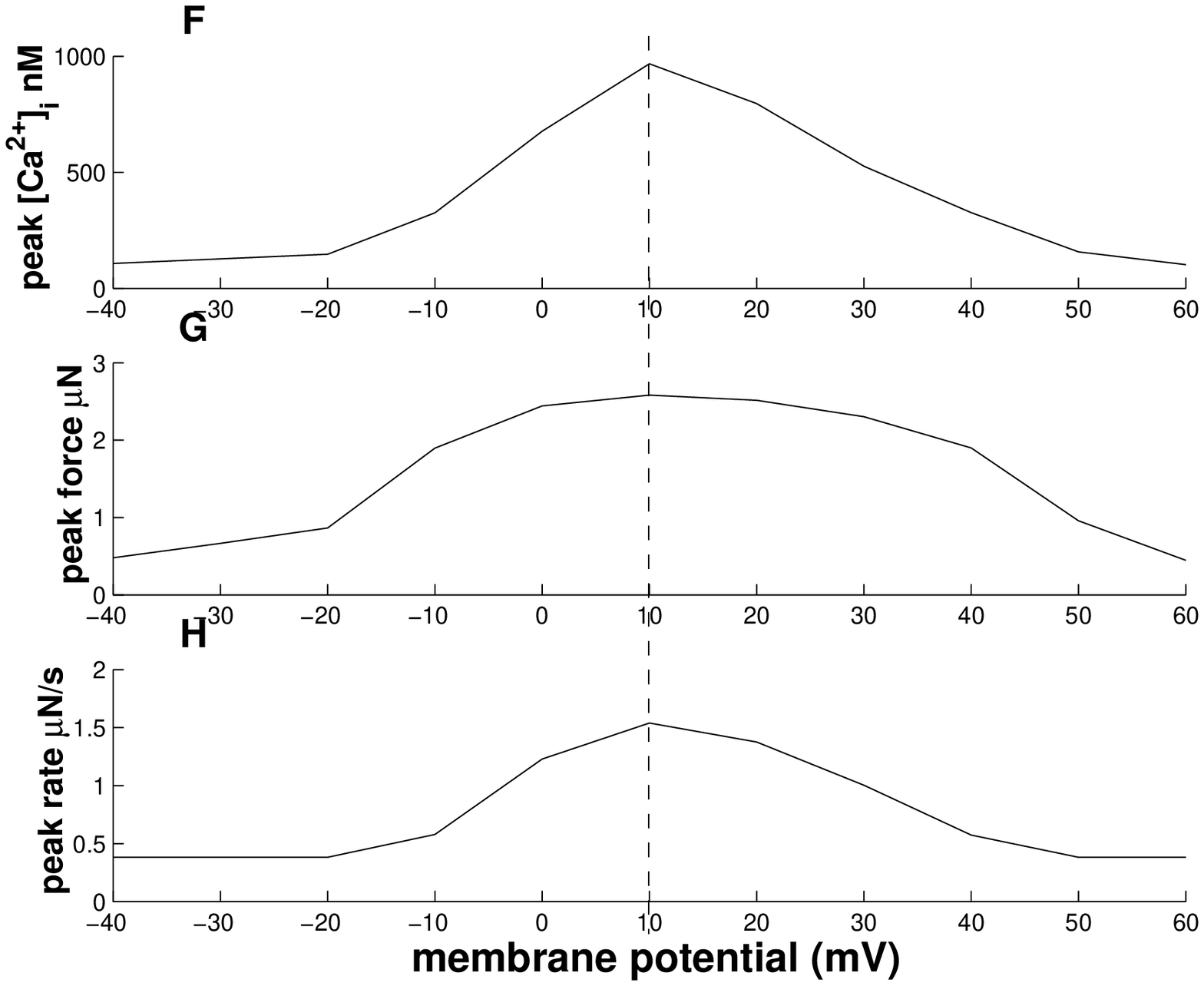}
\caption{\label{fig:whole1} Complete model testing with voltage pulses under isometric condition. A. \Ca \ transient in response to 1.6 s voltage pulse from the holding potential -60 mV to 0 mv. B. The active force developed in the cell in response to the \Ca \ transient. C, D and E. Internal length adjustment of the smooth muscle cell: $l_s$ (series elastic element), $l_a$ (active element) and $l_x$ (cross-bridge element), respectively. $l_{\rm cell}=123$ $\mu$m. F. Peak \Cai under different levels of voltage pulses (from -40 to 60 mV with 10 mV step). G and H. Peak Force and peak rate of force development under different levels of voltage pulses, respectively. The fastest rate of force development corresponds to the largest peak value of \Cai.}
\end{center}
\end{figure}

Figure \ref{fig:whole1} shows a complete model simulation with the cell held under isometric conditions. A depolarizing 1.6s voltage pulse from a holding potential of -60 mV to 0 mV is applied, and the active force generated by smooth muscle contraction follows $[{\rm Ca}^{2+}]$ transient with a delay (Fig.~\ref{fig:whole1}A and B). The voltage pulse is of very short duration relative to the duration of the \Ca \ transient. The model predictions agree well with the measured data of Yagi {\it et al.}~\cite{fay:88a} on single smooth muscle cells. Due to the active cycling of the phosphorylated cross bridges, the length of active element shortens (Fig.~\ref{fig:whole1}D), and correspondingly the lengths of the other two elements ($l_s$ and $l_x$) increase (Panels C and E) to accommodate the increase in force. Under isometric conditions, cell length $L_{\rm cell}$ is constant (123 $\mu$m). {Importantly, Figs.~\ref{fig:whole1}F-H show that peak \Ca \ corresponds directly to the peaks of total force (F) and the rate of force development ($dF/dt$), which is consistent with measured data~\cite{rem:88,fay:88a}.}

\begin{figure}[t]
\begin{center}
\includegraphics[scale=0.45]{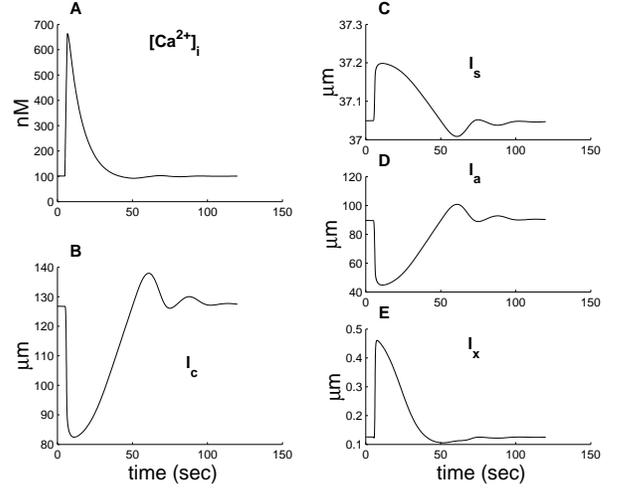}
\caption{\label{fig:whole2} Complete model testing with voltage pulses under isotonic condition. A. \Ca \ transient elicited by 1.6 s voltage pulse from the holding potential -60 mV to 0 mv. B. The cell length change in response to the \Ca \ transient. C, D and E. Internal length adjustments of the cell. Isotonic force: $F=$0.4 $\mu$N.}
\end{center}
\end{figure}

In addition, we investigated the whole cell response under isotonic conditions. In this case, the modeled smooth muscle cell is subjected to the same electrical stimulation protocol delivered under isometric conditions, but rather than a length clamp, a constant force ($F = 0.4$ $\mu$N) is applied to the cell. The electrically elicited \Ca transient is shown in Fig.~\ref{fig:whole2}A, which also resembles the \Ca \ transient shown in Fig.~\ref{fig:whole1}A. Active contraction follows the \Ca \ transient and the change in cell length is shown in Fig.~\ref{fig:whole2}B. Contraction originates in the active force element of the mechanical model, and as the result of the cross-bridge cycling interaction with actin filament, the length of the active element $l_a$ decreases to make contribution to the cell contraction (Fig.~\ref{fig:whole2}D) At the leading edge of this response, the length of the series element and the extension of cross-bridge ($l_s$ and $l_x$, respectively) increase, and because of the imposed isotonic condition, the force in this branch increases to counterbalance the force reduction in the passive branch due to the cell contraction (Figs.~\ref{fig:whole2}C and E). Afterward, the cell begins to relax when the \Ca \ removal mechanisms ($I_{\rm CaP}, I_{\rm up}$) takes over the cytosolic \Ca \ regulation.

\subsubsection{Strain forcing}
\begin{figure}[t]
\centering
\includegraphics[scale=0.38]{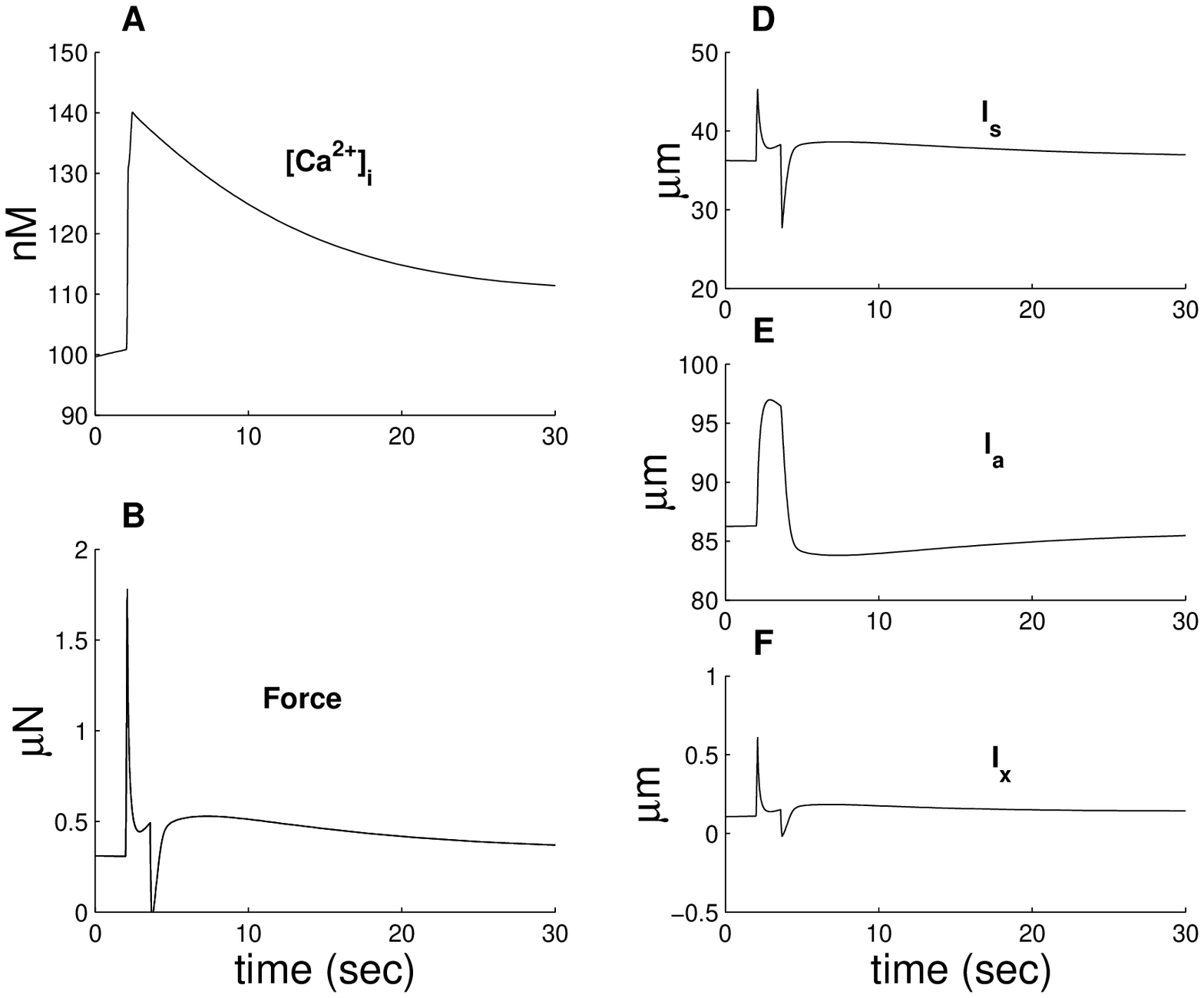}
\includegraphics[scale=0.38]{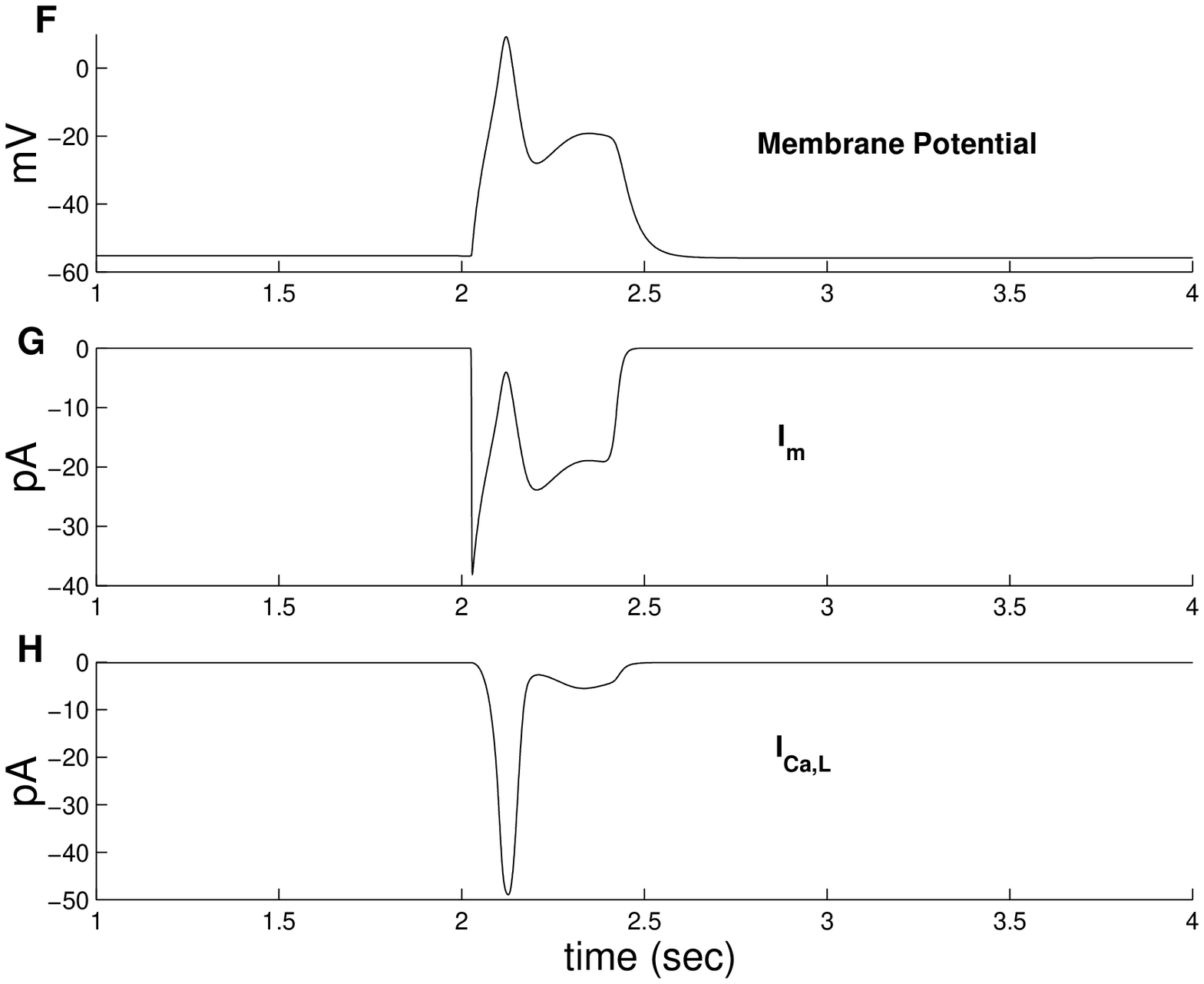}
\caption{\label{fig:whole3} Complete model testing under strain manipulation. 10\% stretch applied to the cell is completed during 100 ms period and is held for 1 second before being brought back to control level (also in 100 ms). A. \Ca \ transient in response to the strain
change. B. Active force (force sustained in active branch) response. C, D and E. Internal length adjustment of the cell. F. Membrane potential in response to the strain change. G and H. Membrane currents elicited by stretch and membrane depolarization ($I_{\rm M}$ and $I_{\rm Ca,L}$, respectively). The control cell length is 123 $\mu$m. Note that the time scale used in Panels F to H is finer than that in Panels A to E.}
\end{figure}

This simulation tests the effect of a strain applied to the smooth muscle cell on the contractile mechanism. Inward current through
the stretch-sensitive channel ($I_{\rm M}$) plays a critical role in depolarizing the membrane, which regulates the voltage-dependent
\Ca \ current $I_{\rm Ca,L}$, and subsequently evokes muscle contraction \cite{dav:99,dav:92a}. A 10\% stretch is applied to the cell as the stimulation, where the strain is defined as the relative change of cell length:
\begin{equation}
\delta_m = \frac{l_{\rm cell} - l_0}{l_0} \ ,
\end{equation}
where $l_0$ is the cell length at which the force is zero. Figs.~\ref{fig:whole3}A-E show the \Ca transient and the mechanical responses. The underlying electrophysiological events are shown in Figs. \ref{fig:whole3}F-H. An initial increase in force occurs immediately after the application of mechanical stretch (Fig.~\ref{fig:whole3}B). It is due to the quick length ($l_s$) increase of the series viscoelastic element (Panel C) and the instantaneous change in length of cross-bridge $l_x$ (Panel E). Force drops quickly due to the sliding mechanism of the active element (Panel D) which reflects the movement of cross-bridges along the actin filament (described as the $\dot{l}_a$ term in (Eq.~\ref{crossfric})). This leads to the relaxation of the spring element on cross-bridges and series element, however, force is still maintained at a lower level until \Ca \ is removed. Membrane depolarization induced by stretch-sensitive current $I_{\rm M}$ (Panel G) activates the L-type \Ca current $I_{\rm Ca,L}$ (Panel H), which contributes to the intracellular \Cai.

\section{Discussion}
We have taken an integrative modeling approach to the characterization of the electrical, chemical, and mechanical behavior of the single cerebrovascular smooth muscle cell. The model takes into account membrane activation by either electrical current pulses or mechanical stretch, and myoplasmic [\Ca] regulation, contractile kinetics of the actin-myosin interaction, force generation and muscle mechanics.

\subsection{Membrane electrophysiology and fluid compartment}
The membrane electrophysiology and fluid compartment models provide very realistic representations of the intracellular \Ca \  transient and hence the input to the contractile model. Characteristics of membrane ionic currents, especially the voltage-dependent calcium current ($I_{\rm Ca,L}$), were determined using voltage-clamp data \cite{rub:96,shi:91} (see Figs.~\ref{figca} and \ref{fig4}). A four-state kinetic model was used to mimic \Ca-induced calcium release (CICR) from the sarcoplasmic reticulum (SR), and measured voltage-clamp induced \Cai \ transients \cite{isen:91,mcn:97} were used to constrain the choice of parameters for the fluid compartment and SR models (Fig.~\ref{catrans}).

\subsection{Myosin phosphorylation and latch bridge}
A combination of hypotheses and measured data were used to form a model of active force generation and mechanical behavior of the cell. We assumed that the Ca$^{2+}\rightarrow$CaCM$\rightarrow$MLCK pathway is very important in myosin phosphorylation and cross-bridge attachment in smooth muscle. A modified version of the multi-state Hai-Murphy (H-M) model~\cite{hai:88a,hai:88b} was employed to describe the kinetics of myosin phosphorylation and cross-bridge attachment (Fig.~\ref{mult1}). Specifically, we used the activated \Ca-calmodulin complex as a trigger for the myosin phosphorylation~\cite{kao6:97}, and made the rate constants $K_1$ and $K_6$ in H-M model functions of CaCM [Eq. (\ref{eq:k16})]. This implies that in our model, the contractile mechanism is tightly coupled with \Ca \ regulation. Other models, however, consider unphosphorylated cross bridge cycling. The cooperative model is an example~\cite{ken:90,som:88,vya:92}, whereby it is assumed that nonphosphorylated cross bridges are activated, attach to actin and cycle under the regulatory influence of a small population of phosphorylated cross bridges. Rembold and Murphy~\cite{mur:93} have evaluated multi-state models of latch bridge formation based on dephosphorylation and/or cooperativity-regulated attachment. They consider several four-state cross-bridge models that have different kinetic schemes, all of which can predict contractile behavior in arterial smooth muscle. Thus, the process of latch bridge formation is uncertain and can be either formed by means of dephosphorylation or cooperativity-regulated attachment~\cite{mur:93}. Due to lack of quantitative data, we have based our model structure on the original kinetic model by Hai and Murphy~\cite{hai:88a,hai:88b}, and have used the same assumption regarding the non-cycling nature of latch-bridges.

\subsection{Cell mechanics}
We have also developed a model of active force generation, and have incorporated it into a mechanical model that describes the mechanical coupling of myofilaments to so called dense region located among the myofilaments and ultimately to the cell wall. Our basic mechanical model of the cell is a modified Hill model, which has a structure that is similar to that developed by Gestrelius {\it et al.} \cite{per:86}. However, it also contains a new description of the coupling between the contractile kinetics and the smooth muscle cell mechanics. Cross-bridge elasticity and mechanical cycling, as well as, the sliding of cross bridges along thin filament are considered, and these processes are modeled as being dependent upon the myosin phosphorylation and cross bridge formation kinetics. The single cell studies of Harris and Warshaw~\cite{dmw:91,dmw:87} provide length:force (L:F) and force:velovity (F:V) data for validation of our mechanical model, and our model provides close fits to the measured data (Figs.~\ref{fig:lf} and \ref{fig:fv}).

\subsection{Model testing and simulations}
Subsystem testing of the integrated model shows: (1) the electrochemical model (membrane-fluid compartment model) not only provides very good fits voltage-clamp data, but to the induced \Ca \ transients as well; and (2) the chemomechnical model (myosin phosphorylation/force development/cell mechanics model) provides close fits to L:F and F:V curves obtained from single smooth muscle cells. The complete integrated cell model was tested by predicting the cell response to both voltage pulse stimulation and strain pulses applied to the cell membrane with the muscle held under isometric conditions (Figs. \ref{fig:whole1} and \ref{fig:whole3}). Additional tests show that the complete cell model can provide simulation of contraction of smooth muscle cell under sustained isotonic tension, to lift a weight W, through a distance (Fig. \ref{fig:whole2}). The component isometric and isotonic phases of this response, mimic closely what is presently
known of smooth muscle mechanics. Our model can also predict contractile events that are difficult to measure, including: (1) the transition occurring between phosphorylated cross-bridges and the latch-bridge state during the response; (2) the internal length adjustments of mechanical elements within the smooth muscle cell, and (3) the integrated responses of several different functional components.

\subsection{Model limitations}
All models have limitations and ours is  no exception.

(1) Our focus in the current study is the development of a model for the single vascular smooth muscle cell. This is a large
task in itself, but the cell model can be integrated into larger models of tissue and vessel in which it resides. Large quantities
of experimental data are available at these more macroscopic levels. The present study should be viewed only as a first step in
the development of an integrative model of the vessel wall, which would necessarily include functional inter-cellular communication
via signaling pathways ({\it e.g.}, nitric oxide, endothelial-smooth muscle gap junctions, etc.). Extensions to tissue and vessel levels
would require additional data regarding the connectivity and inter-cellular communication patterns involved at these levels.

(2) Although experimental studies suggest that a slow inactivation process is responsible for the reduction of the open probability of L-type \Ca \ channels upon membrane depolarization~\cite{rub:96}, quantitative information on the voltage dependent time constant and steady-state of $I_{\rm Ca,L}$ inactivation is lacking. There is no suitable data to quantitatively identify the voltage-dependence of the inactivation kinetics.

(3) The importance of calcium-induced \Ca \ release (CICR) in cerebrovascular smooth muscle is in dispute. Some groups consider
it is important, whereas others disagree. We have provided a model that exhibits CICR by the SR, since CICR may be important in other
types of smooth muscle cell. The peak magnitude of CICR is 10-15\% of the peak of \Ca \ transient (Fig.~\ref{catrans}). Certain
experimental studies suggest that: (a) local \Ca \ release through ryanodine-sensitive SR channels (called ''calcium sparks'') are
responsible for membrane repolarization via a mechanism whereby the released \Ca \ activates \Ca-activated ${\rm K}^+$ channels
($I_{\rm K,Ca}$), which hyperpolarized the cell membrane and limit \Ca \ influx \cite{dav:99,nel:00,nel:98b,nel:95b}. These studies also indicate that CICR does not contribute to the global \Cai \ concentration in cytosol,  and therefore does not directly influence the contractile mechanism via the \Ca-CaCM-MLCK pathway; and (b) IP$_3$-mediated \Ca \ release from SR is alternative pathway for \Cai \ regulation, which has been neglected in the present study, but may need to be considered~\cite{dav:99,htz:96}.

(4) In our model, we assume that \Ca-CaCM-MLCK activated myosin phosphorylation is the major pathway for the contractile kinetics, but alternative pathways may exist. Observations indicate that: (i) activation of protein kinase C induce slow sustained contractions in vascular smooth muscle~\cite{htz:96}, and (ii) Rho-kinase sensitizes the contractile machinery by inhibiting MLCK phosphatase~\cite{scs:01}. Additional data that helps to clarify the relative importance of these various pathways, would be most important to further clarification of our model design.

\section{Acknowledgement} This is work was supported by National Institute of Neurological Disorders and Stroke Grant P01-NS-38660.


\begin{thebibliography}{10}%
\makeatletter
\providecommand \@ifxundefined [1]{%
 \ifx #1\undefined \expandafter \@firstoftwo
 \else \expandafter \@secondoftwo
\fi
}%
\providecommand \@ifnum [1]{%
 \ifnum #1\expandafter \@firstoftwo
 \else \expandafter \@secondoftwo
\fi
}%
\providecommand \enquote [1]{``#1''}%
\providecommand \bibnamefont  [1]{#1}%
\providecommand \bibfnamefont [1]{#1}%
\providecommand \citenamefont [1]{#1}%
\providecommand\href[0]{\@sanitize\@href}%
\providecommand\@href[1]{\endgroup\@@startlink{#1}\endgroup\@@href}%
\providecommand\@@href[1]{#1\@@endlink}%
\providecommand \@sanitize [0]{\begingroup\catcode`\&12\catcode`\#12\relax}%
\@ifxundefined \pdfoutput {\@firstoftwo}{%
 \@ifnum{\z@=\pdfoutput}{\@firstoftwo}{\@secondoftwo}%
}{%
 \providecommand\@@startlink[1]{\leavevmode}%
 \providecommand\@@endlink[0]{}%
}{%
 \providecommand\@@startlink[1]{%
  \leavevmode
  \pdfstartlink
   attr{/Border[0 0 1 ]/H/I/C[0 1 1]}%
   user{/Subtype/Link/A<</Type/Action/S/URI/URI(#1)>>}%
  \relax
 }%
 \providecommand\@@endlink[0]{\pdfendlink}%
}%
\providecommand \url  [0]{\begingroup\@sanitize \@url }%
\providecommand \@url [1]{\endgroup\@href {#1}{\urlprefix}}%
\providecommand \urlprefix [0]{URL }%
\providecommand \Eprint[0]{\href }%
\@ifxundefined \urlstyle {%
  \providecommand \doi [1]{doi:\discretionary{}{}{}#1}%
}{%
  \providecommand \doi [0]{doi:\discretionary{}{}{}\begingroup
  \urlstyle{rm}\Url }%
}%
\providecommand \doibase [0]{http://dx.doi.org/}%
\providecommand \Doi[1]{\href{\doibase#1}}%
\providecommand \bibAnnote [3]{%
  \BibitemShut{#1}%
  \begin{quotation}\noindent
    \textsc{Key:}\ #2\\\textsc{Annotation:}\ #3%
  \end{quotation}%
}%
\providecommand \bibAnnoteFile [2]{%
  \IfFileExists{#2}{\bibAnnote {#1} {#2} {\input{#2}}}{}%
}%
\providecommand \typeout [0]{\immediate \write \m@ne }%
\providecommand \selectlanguage [0]{\@gobble}%
\providecommand \bibinfo [0]{\@secondoftwo}%
\providecommand \bibfield [0]{\@secondoftwo}%
\providecommand \translation [1]{[#1]}%
\providecommand \BibitemOpen[0]{}%
\providecommand \bibitemStop [0]{}%
\providecommand \bibitemNoStop [0]{.\EOS\space}%
\providecommand \EOS [0]{\spacefactor3000\relax}%
\providecommand \BibitemShut [1]{\csname bibitem#1\endcsname}%
\bibitem{lang:96}%
  \BibitemOpen
  \bibfield{author}{%
  \bibinfo {author} {\bibfnamefont{R.~J.}\ \bibnamefont{Lang}}\ and\ \bibinfo
  {author} {\bibfnamefont{C.~A.}\ \bibnamefont{Rattray-Wood}},\ }%
  \emph{\bibinfo {title} {Smooth Muscle Excitation: Chapter 32, A Simple
  Mathematical Model of the Spontaneous Electrical Activity in a Single Smooth
  Muscle Myocyte}}\ (\bibinfo {publisher} {Academic Press},\ \bibinfo {year}
  {1996})%
  \bibAnnoteFile{NoStop}{lang:96}%
\bibitem{wong:93}%
  \BibitemOpen
  \bibfield{author}{%
  \bibinfo {author} {\bibfnamefont{A.~Y.~K.}\ \bibnamefont{Wong}}\ and\
  \bibinfo {author} {\bibfnamefont{G.~A.}\ \bibnamefont{Klassen}},\ }%
  \bibfield{journal}{%
  \bibinfo {journal} {Cell Calcium}\ }%
  \textbf{\bibinfo {volume} {14}},\ \bibinfo {pages} {227} (\bibinfo {year}
  {1993})%
  \bibAnnoteFile{NoStop}{wong:93}%
\bibitem{hai:88a}%
  \BibitemOpen
  \bibfield{author}{%
  \bibinfo {author} {\bibfnamefont{C.}~\bibnamefont{Hai}}\ and\ \bibinfo
  {author} {\bibfnamefont{R.~A.}\ \bibnamefont{Murphy}},\ }%
  \bibfield{journal}{%
  \bibinfo {journal} {Am. J. Physiol.}\ }%
  \textbf{\bibinfo {volume} {254}},\ \bibinfo {pages} {C99} (\bibinfo {year}
  {1988})%
  \bibAnnoteFile{NoStop}{hai:88a}%
\bibitem{hai:88b}%
  \BibitemOpen
  \bibfield{author}{%
  \bibinfo {author} {\bibfnamefont{C.}~\bibnamefont{Hai}}\ and\ \bibinfo
  {author} {\bibfnamefont{R.~A.}\ \bibnamefont{Murphy}},\ }%
  \bibfield{journal}{%
  \bibinfo {journal} {Am. J. Physiol.}\ }%
  \textbf{\bibinfo {volume} {255}},\ \bibinfo {pages} {C86} (\bibinfo {year}
  {1988})%
  \bibAnnoteFile{NoStop}{hai:88b}%
\bibitem{per:86}%
  \BibitemOpen
  \bibfield{author}{%
  \bibinfo {author} {\bibfnamefont{S.}~\bibnamefont{Gestrelius}}\ and\ \bibinfo
  {author} {\bibfnamefont{P.}~\bibnamefont{Borgstr\"{o}m}},\ }%
  \bibfield{journal}{%
  \bibinfo {journal} {Biophysical J.}\ }%
  \textbf{\bibinfo {volume} {50}},\ \bibinfo {pages} {157} (\bibinfo {year}
  {1986})%
  \bibAnnoteFile{NoStop}{per:86}%
\bibitem{dmw:91}%
  \BibitemOpen
  \bibfield{author}{%
  \bibinfo {author} {\bibfnamefont{D.~E.}\ \bibnamefont{Harris}}\ and\ \bibinfo
  {author} {\bibfnamefont{D.~M.}\ \bibnamefont{Warshaw}},\ }%
  \bibfield{journal}{%
  \bibinfo {journal} {Am. J. Physiol.}\ }%
  \textbf{\bibinfo {volume} {260}},\ \bibinfo {pages} {C1104} (\bibinfo {year}
  {1991})%
  \bibAnnoteFile{NoStop}{dmw:91}%
\bibitem{dmw:87}%
  \BibitemOpen
  \bibfield{author}{%
  \bibinfo {author} {\bibfnamefont{D.~M.}\ \bibnamefont{Warshaw}},\ }%
  \bibfield{journal}{%
  \bibinfo {journal} {J. Gen. Physiol.}\ }%
  \textbf{\bibinfo {volume} {89}} (\bibinfo {year} {1987})%
  \bibAnnoteFile{NoStop}{dmw:87}%
\bibitem{fay:83}%
  \BibitemOpen
  \bibfield{author}{%
  \bibinfo {author} {\bibfnamefont{D.~M.}\ \bibnamefont{Warshaw}}\ and\
  \bibinfo {author} {\bibfnamefont{F.~S.}\ \bibnamefont{Fay}},\ }%
  \bibfield{journal}{%
  \bibinfo {journal} {J. Gen. Physiol.}\ }%
  \textbf{\bibinfo {volume} {82}},\ \bibinfo {pages} {157} (\bibinfo {year}
  {1983})%
  \bibAnnoteFile{NoStop}{fay:83}%
\bibitem{fay:88b}%
  \BibitemOpen
  \bibfield{author}{%
  \bibinfo {author} {\bibfnamefont{D.~M.}\ \bibnamefont{Warshaw}}, \bibinfo
  {author} {\bibfnamefont{D.~D.}\ \bibnamefont{Rees}},\ and\ \bibinfo {author}
  {\bibfnamefont{F.~S.}\ \bibnamefont{Fay}},\ }%
  \bibfield{journal}{%
  \bibinfo {journal} {J. Gen. Physiol.}\ }%
  \textbf{\bibinfo {volume} {91}},\ \bibinfo {pages} {761} (\bibinfo {year}
  {1988})%
  \bibAnnoteFile{NoStop}{fay:88b}%
\bibitem{fay:88a}%
  \BibitemOpen
  \bibfield{author}{%
  \bibinfo {author} {\bibfnamefont{S.}~\bibnamefont{Yagi}}, \bibinfo {author}
  {\bibfnamefont{P.~L.}\ \bibnamefont{Becker}},\ and\ \bibinfo {author}
  {\bibfnamefont{F.~S.}\ \bibnamefont{Fay}},\ }%
  \bibfield{journal}{%
  \bibinfo {journal} {Proc. Natl. Acad. Sci. USA.}\ }%
  \textbf{\bibinfo {volume} {85}},\ \bibinfo {pages} {4109} (\bibinfo {year}
  {1988})%
  \bibAnnoteFile{NoStop}{fay:88a}%
\bibitem{rub:96}%
  \BibitemOpen
  \bibfield{author}{%
  \bibinfo {author} {\bibfnamefont{M.}~\bibnamefont{Rubart}}, \bibinfo {author}
  {\bibfnamefont{J.~B.}\ \bibnamefont{Patlak}},\ and\ \bibinfo {author}
  {\bibfnamefont{M.~T.}\ \bibnamefont{Nelson}},\ }%
  \bibfield{journal}{%
  \bibinfo {journal} {J. Gen. Physiol.}\ }%
  \textbf{\bibinfo {volume} {107}},\ \bibinfo {pages} {459} (\bibinfo {year}
  {1996})%
  \bibAnnoteFile{NoStop}{rub:96}%
\bibitem{lan:93a}%
  \BibitemOpen
  \bibfield{author}{%
  \bibinfo {author} {\bibfnamefont{P.~D.}\ \bibnamefont{Langton}},\ }%
  \bibfield{journal}{%
  \bibinfo {journal} {J. Physiol. (Lond.)}\ }%
  \textbf{\bibinfo {volume} {471}},\ \bibinfo {pages} {1} (\bibinfo {year}
  {1993})%
  \bibAnnoteFile{NoStop}{lan:93a}%
\bibitem{dav:99}%
  \BibitemOpen
  \bibfield{author}{%
  \bibinfo {author} {\bibfnamefont{M.~J.}\ \bibnamefont{Davis}}\ and\ \bibinfo
  {author} {\bibfnamefont{M.~A.}\ \bibnamefont{Hill}},\ }%
  \bibfield{journal}{%
  \bibinfo {journal} {Physiol. Rev.}\ }%
  \textbf{\bibinfo {volume} {79}},\ \bibinfo {pages} {387} (\bibinfo {year}
  {1999})%
  \bibAnnoteFile{NoStop}{dav:99}%
\bibitem{htz:96}%
  \BibitemOpen
  \bibfield{author}{%
  \bibinfo {author} {\bibfnamefont{A.}~\bibnamefont{Horowitz}}, \bibinfo
  {author} {\bibfnamefont{C.~B.}\ \bibnamefont{Menice}}, \bibinfo {author}
  {\bibfnamefont{R.}~\bibnamefont{Laporte}},\ and\ \bibinfo {author}
  {\bibfnamefont{K.~G.}\ \bibnamefont{Morgan}},\ }%
  \bibfield{journal}{%
  \bibinfo {journal} {Physiol. Rev.}\ }%
  \textbf{\bibinfo {volume} {76}},\ \bibinfo {pages} {967} (\bibinfo {year}
  {1996})%
  \bibAnnoteFile{NoStop}{htz:96}%
\bibitem{jrm:98}%
  \BibitemOpen
  \bibfield{author}{%
  \bibinfo {author} {\bibfnamefont{J.~R.}\ \bibnamefont{Martens}}\ and\
  \bibinfo {author} {\bibfnamefont{C.~H.}\ \bibnamefont{Gelband}},\ }%
  \bibfield{journal}{%
  \bibinfo {journal} {P.S.E.B.M}\ }%
  \textbf{\bibinfo {volume} {218}},\ \bibinfo {pages} {192} (\bibinfo {year}
  {1998})%
  \bibAnnoteFile{NoStop}{jrm:98}%
\bibitem{dav:92a}%
  \BibitemOpen
  \bibfield{author}{%
  \bibinfo {author} {\bibfnamefont{G.~A.}\ \bibnamefont{Meininger}}\ and\
  \bibinfo {author} {\bibfnamefont{M.~J.}\ \bibnamefont{Davis}},\ }%
  \bibfield{journal}{%
  \bibinfo {journal} {Am. J. Physiol.}\ }%
  \textbf{\bibinfo {volume} {263}},\ \bibinfo {pages} {H647} (\bibinfo {year}
  {1992})%
  \bibAnnoteFile{NoStop}{dav:92a}%
\bibitem{nel:90}%
  \BibitemOpen
  \bibfield{author}{%
  \bibinfo {author} {\bibfnamefont{M.~T.}\ \bibnamefont{Nelson}}, \bibinfo
  {author} {\bibfnamefont{J.~B.}\ \bibnamefont{Patlak}}, \bibinfo {author}
  {\bibfnamefont{J.~F.}\ \bibnamefont{Worley}},\ and\ \bibinfo {author}
  {\bibfnamefont{N.~B.}\ \bibnamefont{Standen}},\ }%
  \bibfield{journal}{%
  \bibinfo {journal} {Am. J. Physiol.}\ }%
  \textbf{\bibinfo {volume} {259}},\ \bibinfo {pages} {C3} (\bibinfo {year}
  {1990})%
  \bibAnnoteFile{NoStop}{nel:90}%
\bibitem{adh:95}%
  \BibitemOpen
  \bibfield{author}{%
  \bibinfo {author} {\bibfnamefont{A.~D.}\ \bibnamefont{Hughes}},\ }%
  \bibfield{journal}{%
  \bibinfo {journal} {J. Vascular Res.}\ }%
  \textbf{\bibinfo {volume} {32}},\ \bibinfo {pages} {353} (\bibinfo {year}
  {1995})%
  \bibAnnoteFile{NoStop}{adh:95}%
\bibitem{nrc:93}%
  \BibitemOpen
  \bibfield{author}{%
  \bibinfo {author} {\bibfnamefont{W.~H.}\ \bibnamefont{Press}}, \bibinfo
  {author} {\bibfnamefont{S.~A.}\ \bibnamefont{Teukolsky}}, \bibinfo {author}
  {\bibfnamefont{W.~T.}\ \bibnamefont{Vetterling}},\ and\ \bibinfo {author}
  {\bibfnamefont{B.~P.}\ \bibnamefont{Flannery}},\ }%
  \emph{\bibinfo {title} {Numerical Recipes in C: The Art of Scientific
  Computing}},\ \bibinfo {edition} {2nd}\ ed.\ (\bibinfo {publisher} {Cambridge
  University Press},\ \bibinfo {year} {1993})%
  \bibAnnoteFile{NoStop}{nrc:93}%
\bibitem{dav:92b}%
  \BibitemOpen
  \bibfield{author}{%
  \bibinfo {author} {\bibfnamefont{M.~J.}\ \bibnamefont{Davis}}, \bibinfo
  {author} {\bibfnamefont{J.~A.}\ \bibnamefont{Donovitz}},\ and\ \bibinfo
  {author} {\bibfnamefont{J.~D.}\ \bibnamefont{Hood}},\ }%
  \bibfield{journal}{%
  \bibinfo {journal} {Am. J. Physiol.}\ }%
  \textbf{\bibinfo {volume} {262}},\ \bibinfo {pages} {C1083} (\bibinfo {year}
  {1992})%
  \bibAnnoteFile{NoStop}{dav:92b}%
\bibitem{shi:91}%
  \BibitemOpen
  \bibfield{author}{%
  \bibinfo {author} {\bibfnamefont{K.~A.}\ \bibnamefont{Volk}}, \bibinfo
  {author} {\bibfnamefont{J.~J.}\ \bibnamefont{Matsuda}},\ and\ \bibinfo
  {author} {\bibfnamefont{E.~F.}\ \bibnamefont{Shibata}},\ }%
  \bibfield{journal}{%
  \bibinfo {journal} {J. Physiol. (Lond.)}\ }%
  \textbf{\bibinfo {volume} {439}},\ \bibinfo {pages} {751} (\bibinfo {year}
  {1991})%
  \bibAnnoteFile{NoStop}{shi:91}%
\bibitem{nel:94}%
  \BibitemOpen
  \bibfield{author}{%
  \bibinfo {author} {\bibfnamefont{B.~E.}\ \bibnamefont{Roberson}}\ and\
  \bibinfo {author} {\bibfnamefont{M.~T.}\ \bibnamefont{Nelson}},\ }%
  \bibfield{journal}{%
  \bibinfo {journal} {Am. J. Physiol.}\ }%
  \textbf{\bibinfo {volume} {267}},\ \bibinfo {pages} {C1589} (\bibinfo {year}
  {1994})%
  \bibAnnoteFile{NoStop}{nel:94}%
\bibitem{nel:95}%
  \BibitemOpen
  \bibfield{author}{%
  \bibinfo {author} {\bibfnamefont{M.~T.}\ \bibnamefont{Nelson}}\ and\ \bibinfo
  {author} {\bibfnamefont{J.~M.}\ \bibnamefont{Quayle}},\ }%
  \bibfield{journal}{%
  \bibinfo {journal} {Am. J. Physiol.}\ }%
  \textbf{\bibinfo {volume} {268}},\ \bibinfo {pages} {C799} (\bibinfo {year}
  {1995})%
  \bibAnnoteFile{NoStop}{nel:95}%
\bibitem{wang:93}%
  \BibitemOpen
  \bibfield{author}{%
  \bibinfo {author} {\bibfnamefont{Y.}~\bibnamefont{Wang}}\ and\ \bibinfo
  {author} {\bibfnamefont{D.~A.}\ \bibnamefont{Mathers}},\ }%
  \bibfield{journal}{%
  \bibinfo {journal} {J. Physiol. (Lond.)}\ }%
  \textbf{\bibinfo {volume} {462}},\ \bibinfo {pages} {529} (\bibinfo {year}
  {1993})%
  \bibAnnoteFile{NoStop}{wang:93}%
\bibitem{hirst:88}%
  \BibitemOpen
  \bibfield{author}{%
  \bibinfo {author} {\bibfnamefont{F.~R.}\ \bibnamefont{Edwards}}, \bibinfo
  {author} {\bibfnamefont{G.~D.~S.}\ \bibnamefont{Hirst}},\ and\ \bibinfo
  {author} {\bibfnamefont{G.~D.}\ \bibnamefont{Silverberg}},\ }%
  \bibfield{journal}{%
  \bibinfo {journal} {J. Physiol. (Lond.)}\ }%
  \textbf{\bibinfo {volume} {404}},\ \bibinfo {pages} {455} (\bibinfo {year}
  {1988})%
  \bibAnnoteFile{NoStop}{hirst:88}%
\bibitem{jmq:96}%
  \BibitemOpen
  \bibfield{author}{%
  \bibinfo {author} {\bibfnamefont{J.~M.}\ \bibnamefont{Quayle}}, \bibinfo
  {author} {\bibfnamefont{C.}~\bibnamefont{Dart}},\ and\ \bibinfo {author}
  {\bibfnamefont{N.~B.}\ \bibnamefont{Standen}},\ }%
  \bibfield{journal}{%
  \bibinfo {journal} {J. Physiol. (Lond.)}\ }%
  \textbf{\bibinfo {volume} {494}},\ \bibinfo {pages} {715} (\bibinfo {year}
  {1996})%
  \bibAnnoteFile{NoStop}{jmq:96}%
\bibitem{jwc:96}%
  \BibitemOpen
  \bibfield{author}{%
  \bibinfo {author} {\bibfnamefont{D.~S.}\ \bibnamefont{Lindblad}}, \bibinfo
  {author} {\bibfnamefont{C.~R.}\ \bibnamefont{Murphy}}, \bibinfo {author}
  {\bibfnamefont{J.~W.}\ \bibnamefont{Clark}},\ and\ \bibinfo {author}
  {\bibfnamefont{W.~R.}\ \bibnamefont{Giles}},\ }%
  \bibfield{journal}{%
  \bibinfo {journal} {Am. J. Physiol.}\ }%
  \textbf{\bibinfo {volume} {271}},\ \bibinfo {pages} {H1666} (\bibinfo {year}
  {1996})%
  \bibAnnoteFile{NoStop}{jwc:96}%
\bibitem{meo:94}%
  \BibitemOpen
  \bibfield{author}{%
  \bibinfo {author} {\bibfnamefont{M.~E.}\ \bibnamefont{O'Donnell}}\ and\
  \bibinfo {author} {\bibfnamefont{N.~E.}\ \bibnamefont{Owen}},\ }%
  \bibfield{journal}{%
  \bibinfo {journal} {Physiol. Rev.}\ }%
  \textbf{\bibinfo {volume} {74}},\ \bibinfo {pages} {683} (\bibinfo {year}
  {1994})%
  \bibAnnoteFile{NoStop}{meo:94}%
\bibitem{mcn:94}%
  \BibitemOpen
  \bibfield{author}{%
  \bibinfo {author} {\bibfnamefont{J.~G.}\ \bibnamefont{McCarron}}\ and\
  \bibinfo {author} {\bibfnamefont{J.~V.~W.}\ \bibnamefont{nd~F.~S.~Fay}},\ }%
  \bibfield{journal}{%
  \bibinfo {journal} {Pfl\"{u}gers Arch.}\ }%
  \textbf{\bibinfo {volume} {426}},\ \bibinfo {pages} {199} (\bibinfo {year}
  {1994})%
  \bibAnnoteFile{NoStop}{mcn:94}%
\bibitem{isen:91}%
  \BibitemOpen
  \bibfield{author}{%
  \bibinfo {author} {\bibfnamefont{V.~Y.}\ \bibnamefont{Ganitkevich}}\ and\
  \bibinfo {author} {\bibfnamefont{G.}~\bibnamefont{Isenberg}},\ }%
  \bibfield{journal}{%
  \bibinfo {journal} {J. Physiol. (Lond.)}\ }%
  \textbf{\bibinfo {volume} {435}},\ \bibinfo {pages} {187} (\bibinfo {year}
  {1991})%
  \bibAnnoteFile{NoStop}{isen:91}%
\bibitem{mcn:98}%
  \BibitemOpen
  \bibfield{author}{%
  \bibinfo {author} {\bibfnamefont{T.}~\bibnamefont{Kamishima}}\ and\ \bibinfo
  {author} {\bibfnamefont{J.~G.}\ \bibnamefont{McCarron}},\ }%
  \bibfield{journal}{%
  \bibinfo {journal} {Biophysical J.}\ }%
  \textbf{\bibinfo {volume} {75}},\ \bibinfo {pages} {1767} (\bibinfo {year}
  {1998})%
  \bibAnnoteFile{NoStop}{mcn:98}%
\bibitem{nel:00}%
  \BibitemOpen
  \bibfield{author}{%
  \bibinfo {author} {\bibfnamefont{J.~H.}\ \bibnamefont{Jagger}}, \bibinfo
  {author} {\bibfnamefont{V.~A.}\ \bibnamefont{Porter}}, \bibinfo {author}
  {\bibfnamefont{W.~J.}\ \bibnamefont{Lederer}},\ and\ \bibinfo {author}
  {\bibfnamefont{M.~T.}\ \bibnamefont{Nelson}},\ }%
  \bibfield{journal}{%
  \bibinfo {journal} {Am. J. Physiol.}\ }%
  \textbf{\bibinfo {volume} {278}},\ \bibinfo {pages} {C235} (\bibinfo {year}
  {2000})%
  \bibAnnoteFile{NoStop}{nel:00}%
\bibitem{nel:98b}%
  \BibitemOpen
  \bibfield{author}{%
  \bibinfo {author} {\bibfnamefont{H.~J.}\ \bibnamefont{Knot}}, \bibinfo
  {author} {\bibfnamefont{N.~B.}\ \bibnamefont{Standen}},\ and\ \bibinfo
  {author} {\bibfnamefont{M.~T.}\ \bibnamefont{Nelson}},\ }%
  \bibfield{journal}{%
  \bibinfo {journal} {J. Physiol. (Lond.)}\ }%
  \textbf{\bibinfo {volume} {508}},\ \bibinfo {pages} {211} (\bibinfo {year}
  {1998})%
  \bibAnnoteFile{NoStop}{nel:98b}%
\bibitem{mcn:97}%
  \BibitemOpen
  \bibfield{author}{%
  \bibinfo {author} {\bibfnamefont{T.}~\bibnamefont{Kamishima}}\ and\ \bibinfo
  {author} {\bibfnamefont{J.~G.}\ \bibnamefont{McCarron}},\ }%
  \bibfield{journal}{%
  \bibinfo {journal} {J. Physiol. (Lond.)}\ }%
  \textbf{\bibinfo {volume} {501}},\ \bibinfo {pages} {497} (\bibinfo {year}
  {1997})%
  \bibAnnoteFile{NoStop}{mcn:97}%
\bibitem{fab:92}%
  \BibitemOpen
  \bibfield{author}{%
  \bibinfo {author} {\bibfnamefont{A.}~\bibnamefont{Fabiato}},\ }%
  \emph{\bibinfo {title} {Excitation-contraction coupling in skeletal, cardiac,
  and smooth muscle: Two kinds of calcium--induced release of calcium from
  sarcoplasmic reticulum of skinned cardiac cells}}\ (\bibinfo {publisher}
  {Plenum Press},\ \bibinfo {year} {1992})%
  \bibAnnoteFile{NoStop}{fab:92}%
\bibitem{stn:99}%
  \BibitemOpen
  \bibfield{author}{%
  \bibinfo {author} {\bibfnamefont{M.~D.}\ \bibnamefont{Stern}}, \bibinfo
  {author} {\bibfnamefont{L.}~\bibnamefont{Song}}, \bibinfo {author}
  {\bibfnamefont{H.}~\bibnamefont{Cheng}}, \bibinfo {author}
  {\bibfnamefont{J.~S.~K.}\ \bibnamefont{Shaw}}, \bibinfo {author}
  {\bibfnamefont{H.~T.}\ \bibnamefont{Yang}}, \bibinfo {author}
  {\bibfnamefont{K.~R.}\ \bibnamefont{Boheler}},\ and\ \bibinfo {author}
  {\bibfnamefont{E.}~\bibnamefont{Rios}},\ }%
  \bibfield{journal}{%
  \bibinfo {journal} {J. Gen. Physiol.}\ }%
  \textbf{\bibinfo {volume} {113}},\ \bibinfo {pages} {469} (\bibinfo {year}
  {1999})%
  \bibAnnoteFile{NoStop}{stn:99}%
\bibitem{tang:94}%
  \BibitemOpen
  \bibfield{author}{%
  \bibinfo {author} {\bibfnamefont{Y.}~\bibnamefont{Tang}}\ and\ \bibinfo
  {author} {\bibfnamefont{H.~G.}\ \bibnamefont{Othmer}},\ }%
  \bibfield{journal}{%
  \bibinfo {journal} {Biophysical J.}\ }%
  \textbf{\bibinfo {volume} {67}},\ \bibinfo {pages} {2223} (\bibinfo {year}
  {1994})%
  \bibAnnoteFile{NoStop}{tang:94}%
\bibitem{kao2:97}%
  \BibitemOpen
  \bibfield{author}{%
  \bibinfo {author} {\bibfnamefont{J.~D.}\ \bibnamefont{Miller}}\ and\ \bibinfo
  {author} {\bibfnamefont{M.~E.}\ \bibnamefont{Carsten}},\ }%
  \emph{\bibinfo {title} {Cellular aspects of smooth muscle function: Chatper
  2, Calcium homeostasis in smooth muscle}}\ (\bibinfo {publisher} {Camgridge
  University Press},\ \bibinfo {year} {1997})%
  \bibAnnoteFile{NoStop}{kao2:97}%
\bibitem{wang:95}%
  \BibitemOpen
  \bibfield{author}{%
  \bibinfo {author} {\bibfnamefont{C.-L.~A.}\ \bibnamefont{Wang}},\ }%
  \bibfield{journal}{%
  \bibinfo {journal} {Biochem. Bioohys. Res. Commun.}\ }%
  \textbf{\bibinfo {volume} {130}},\ \bibinfo {pages} {426} (\bibinfo {year}
  {1995})%
  \bibAnnoteFile{NoStop}{wang:95}%
\bibitem{mur:90}%
  \BibitemOpen
  \bibfield{author}{%
  \bibinfo {author} {\bibfnamefont{R.~A.}\ \bibnamefont{Murphy}}, \bibinfo
  {author} {\bibfnamefont{C.~M.}\ \bibnamefont{Rembold}},\ and\ \bibinfo
  {author} {\bibfnamefont{C.}~\bibnamefont{Hai}},\ }%
  \bibfield{journal}{%
  \bibinfo {journal} {Prog. Clin. Biol. Res.}\ }%
  \textbf{\bibinfo {volume} {327}},\ \bibinfo {pages} {39} (\bibinfo {year}
  {1990})%
  \bibAnnoteFile{NoStop}{mur:90}%
\bibitem{bar:96}%
  \BibitemOpen
  \emph{\bibinfo {title} {Biochemistry of Smooth Muscle Contraction}},\ edited
  by\ \bibinfo {editor} {\bibfnamefont{M.}~\bibnamefont{B\'{a}r\'{a}ny}}\
  (\bibinfo {publisher} {Academic Press},\ \bibinfo {year} {1996})%
  \bibAnnoteFile{NoStop}{bar:96}%
\bibitem{ycf:93}%
  \BibitemOpen
  \bibfield{author}{%
  \bibinfo {author} {\bibfnamefont{Y.~C.}\ \bibnamefont{Fung}},\ }%
  \emph{\bibinfo {title} {Biomechnics: Mechanical Properties of Living
  Tissues}},\ \bibinfo {edition} {2nd}\ ed.\ (\bibinfo {publisher}
  {Springer--Verlag},\ \bibinfo {year} {1993})%
  \bibAnnoteFile{NoStop}{ycf:93}%
\bibitem{mur:83}%
  \BibitemOpen
  \bibfield{author}{%
  \bibinfo {author} {\bibfnamefont{M.~O.}\ \bibnamefont{Aksoy}}, \bibinfo
  {author} {\bibfnamefont{S.}~\bibnamefont{Mras}}, \bibinfo {author}
  {\bibfnamefont{K.~E.}\ \bibnamefont{Kamm}},\ and\ \bibinfo {author}
  {\bibfnamefont{R.~A.}\ \bibnamefont{Murphy}},\ }%
  \bibfield{journal}{%
  \bibinfo {journal} {Am. J. Physiol.}\ }%
  \textbf{\bibinfo {volume} {245}},\ \bibinfo {pages} {C255} (\bibinfo {year}
  {1983})%
  \bibAnnoteFile{NoStop}{mur:83}%
\bibitem{dmw:90}%
  \BibitemOpen
  \bibfield{author}{%
  \bibinfo {author} {\bibfnamefont{D.~M.}\ \bibnamefont{Warshaw}}, \bibinfo
  {author} {\bibfnamefont{J.~M.}\ \bibnamefont{Desrosiers}}, \bibinfo {author}
  {\bibfnamefont{S.~S.}\ \bibnamefont{Work}},\ and\ \bibinfo {author}
  {\bibfnamefont{K.~M.}\ \bibnamefont{Trybus}},\ }%
  \bibfield{journal}{%
  \bibinfo {journal} {J. Cell Biol.}\ }%
  \textbf{\bibinfo {volume} {111}},\ \bibinfo {pages} {453} (\bibinfo {year}
  {1990})%
  \bibAnnoteFile{NoStop}{dmw:90}%
\bibitem{hill:38}%
  \BibitemOpen
  \bibfield{author}{%
  \bibinfo {author} {\bibfnamefont{A.~V.}\ \bibnamefont{Hill}},\ }%
  \bibfield{journal}{%
  \bibinfo {journal} {Proc. Roy. Soc. Lond. B. Biol. Sci.}\ }%
  \textbf{\bibinfo {volume} {126}},\ \bibinfo {pages} {136} (\bibinfo {year}
  {1938})%
  \bibAnnoteFile{NoStop}{hill:38}%
\bibitem{rem:88}%
  \BibitemOpen
  \bibfield{author}{%
  \bibinfo {author} {\bibfnamefont{C.~M.}\ \bibnamefont{Rembold}}\ and\
  \bibinfo {author} {\bibfnamefont{R.~A.}\ \bibnamefont{Murphy}},\ }%
  \bibfield{journal}{%
  \bibinfo {journal} {J. Cardiovasc. Pharmacol.}\ }%
  \textbf{\bibinfo {volume} {12}},\ \bibinfo {pages} {S38} (\bibinfo {year}
  {1988})%
  \bibAnnoteFile{NoStop}{rem:88}%
\bibitem{kao6:97}%
  \BibitemOpen
  \bibfield{author}{%
  \bibinfo {author} {\bibfnamefont{R.~A.}\ \bibnamefont{Word}}\ and\ \bibinfo
  {author} {\bibfnamefont{K.~E.}\ \bibnamefont{Kamm}},\ }%
  \emph{\bibinfo {title} {Cellular aspects of smooth muscle function: Chapter
  6, Regulation of smooth muscle contraction by myosin phosphorylation}}\
  (\bibinfo {publisher} {Camgridge University Press},\ \bibinfo {year} {1997})%
  \bibAnnoteFile{NoStop}{kao6:97}%
\bibitem{ken:90}%
  \BibitemOpen
  \bibfield{author}{%
  \bibinfo {author} {\bibfnamefont{R.~E.}\ \bibnamefont{Kenney}}, \bibinfo
  {author} {\bibfnamefont{P.~E.}\ \bibnamefont{Hoar}},\ and\ \bibinfo {author}
  {\bibfnamefont{W.~G.~L.}\ \bibnamefont{Kerrick}},\ }%
  \bibfield{journal}{%
  \bibinfo {journal} {J. Biol. Chem.}\ }%
  \textbf{\bibinfo {volume} {265}},\ \bibinfo {pages} {8642} (\bibinfo {year}
  {1990})%
  \bibAnnoteFile{NoStop}{ken:90}%
\bibitem{som:88}%
  \BibitemOpen
  \bibfield{author}{%
  \bibinfo {author} {\bibfnamefont{A.~V.}\ \bibnamefont{Somlyo}}, \bibinfo
  {author} {\bibfnamefont{Y.~E.}\ \bibnamefont{Goldman}}, \bibinfo {author}
  {\bibfnamefont{T.}~\bibnamefont{Fujimori}}, \bibinfo {author}
  {\bibfnamefont{M.}~\bibnamefont{Bond}}, \bibinfo {author}
  {\bibfnamefont{D.~R.}\ \bibnamefont{Trentham}},\ and\ \bibinfo {author}
  {\bibfnamefont{A.~P.}\ \bibnamefont{Somlyo}},\ }%
  \bibfield{journal}{%
  \bibinfo {journal} {J. Gen. Physiol.}\ }%
  \textbf{\bibinfo {volume} {91}},\ \bibinfo {pages} {165} (\bibinfo {year}
  {1988})%
  \bibAnnoteFile{NoStop}{som:88}%
\bibitem{vya:92}%
  \BibitemOpen
  \bibfield{author}{%
  \bibinfo {author} {\bibfnamefont{T.~B.}\ \bibnamefont{Vyas}}, \bibinfo
  {author} {\bibfnamefont{S.~U.}\ \bibnamefont{Mooers}}, \bibinfo {author}
  {\bibfnamefont{S.~R.}\ \bibnamefont{Narayan}}, \bibinfo {author}
  {\bibfnamefont{J.~C.}\ \bibnamefont{Witherell}}, \bibinfo {author}
  {\bibfnamefont{M.~J.}\ \bibnamefont{Siegman}},\ and\ \bibinfo {author}
  {\bibfnamefont{T.~M.}\ \bibnamefont{Butler}},\ }%
  \bibfield{journal}{%
  \bibinfo {journal} {Am. J. Physiol.}\ }%
  \textbf{\bibinfo {volume} {263}},\ \bibinfo {pages} {C210} (\bibinfo {year}
  {1992})%
  \bibAnnoteFile{NoStop}{vya:92}%
\bibitem{mur:93}%
  \BibitemOpen
  \bibfield{author}{%
  \bibinfo {author} {\bibfnamefont{C.~M.}\ \bibnamefont{Rembold}}\ and\
  \bibinfo {author} {\bibfnamefont{R.~A.}\ \bibnamefont{Murphy}},\ }%
  \bibfield{journal}{%
  \bibinfo {journal} {J. Muscle Res. Cell. Motil.}\ }%
  \textbf{\bibinfo {volume} {14}},\ \bibinfo {pages} {325} (\bibinfo {year}
  {1993})%
  \bibAnnoteFile{NoStop}{mur:93}%
\bibitem{nel:95b}%
  \BibitemOpen
  \bibfield{author}{%
  \bibinfo {author} {\bibfnamefont{M.~T.}\ \bibnamefont{Nelson}}, \bibinfo
  {author} {\bibfnamefont{H.}~\bibnamefont{Cheng}}, \bibinfo {author}
  {\bibfnamefont{M.}~\bibnamefont{Rubart}}, \bibinfo {author}
  {\bibfnamefont{L.~F.}\ \bibnamefont{Santana}}, \bibinfo {author}
  {\bibfnamefont{A.~D.}\ \bibnamefont{Bonev}}, \bibinfo {author}
  {\bibfnamefont{H.~J.}\ \bibnamefont{Kont}},\ and\ \bibinfo {author}
  {\bibfnamefont{W.~J.}\ \bibnamefont{Lederer}},\ }%
  \bibfield{journal}{%
  \bibinfo {journal} {Science}\ }%
  \textbf{\bibinfo {volume} {270}},\ \bibinfo {pages} {633} (\bibinfo {year}
  {1995})%
  \bibAnnoteFile{NoStop}{nel:95b}%
\bibitem{scs:01}%
  \BibitemOpen
  \bibfield{author}{%
  \bibinfo {author} {\bibfnamefont{S.}~\bibnamefont{Chrissobolis}}\ and\
  \bibinfo {author} {\bibfnamefont{C.~G.}\ \bibnamefont{Sobey}},\ }%
  \bibfield{journal}{%
  \bibinfo {journal} {Circ. Res.}\ }%
  \textbf{\bibinfo {volume} {88}},\ \bibinfo {pages} {774} (\bibinfo {year}
  {2001})%
  \bibAnnoteFile{NoStop}{scs:01}%
\end{thebibliography}
%

\section{APPENDIX - Mathematical equations for the electrochemical model}\label{appx1}
Mathematical equations for the model and related parameter values are listed in this appendix.

\clearpage
\begin{table*}[t]
\begin{center}
\caption{\label{tabpara1} Parameters for electro-chemical model}
\begin{tabular}{|l|l|l|}\hline
Parameter & Description & Numerical value \\
\hline\hline
R & ideal gas constant & 8341.0 mJ$\cdot$mol$^{-1}$K$^{-1}$\\
T & absolute temperature & 293 $K$ \\
F & Faraday's constant & 96487.0 C$\cdot$mol$^{-1}$\\
$C_m$ & membrane capacitance & 0.03 nF \\
$g_{\rm b,Na}$ & maximum background ${\rm Na}^+$ current conductance & 0.01 nS \\
$g_{\rm b,Ca}$ & maximum background \Ca current conductance & 0.012 nS \\
$g_{\rm b,K}$ & maximum background ${\rm K}^+$ current conductance & 0.01 nS \\
$G_{\rm K_i}$ & inward rectifier constant & 0.145 \\
$n_{\rm K_i}$ &inward rectifier constant & 0.5 \\
$g_{\rm KCa}$ & maximum \Ca-activated ${\rm K}^+$ current conductance & 0.5 nS \\
$V_{\rm 1/2,KCa}$ & half-activation potential for $I_{\rm K,Ca}$ & 20.5 mV \\
$\bar{I}_{\rm CaP}$ & maximum \Ca pump current & 8.3 pA\\
$\bar{I}_{\rm NaK}$ & maximum ${\rm Na}^+$-${\rm K}^+$ current & 7.5 pA \\
$K_{\rm m,K}$ & concentration of half-activation for $K^+$ & 1.0 mM\\
$K_{\rm m,Na}$ & concentration of half-activation for ${\rm Na}^+$ & 11.0 mM\\
$k_{\rm NaCa}$ & intrinsic current density of ${\rm Na}^+$-\Ca \ exchanger & 0.0005 pA mM$^{-4}$ \\
$d_{\rm NaCa}$ & scaling factor for ${\rm Na}^+$-\Ca \ exchanger & 0.0003 mM$^{-4}$ \\
$\gamma$ & partition coefficient for ${\rm Na}^+$-\Ca exchanger & 0.45 \\
${\rm vol}_i$ & volume of intracellular space & 1.0 pl \\
${\rm vol}_{\rm Ca}$ & intracellular volume available to free \Ca & 0.7 pl \\
$[{\rm Ca}^{2+}]_o$ & extracellular calcium concentration & 2.0 mM \\
$[{\rm Na}^{+}]_o$ & extracellular sodium concentration & 140.0 mM \\
$[{\rm K}^{+}]_o$ & extracellular potassium concentration & 5.0 mM \\
$[\bar{S}_{\rm CM}]$ & total concentration of calmodulin sites for \Ca & 0.1 mM \\
$[\bar{B}_F]$ & total concentration of other buffer sites for \Ca & 0.4 mM \\
$k_1$ & \Ca-calmodulin association rate constant & 200 mM$^{-1}s^{-1}$\\
$k_{-1}$ & \Ca-calmodulin dissociation rate constant & 0.052 s$^{-1}$\\
$k_d$ & \Ca-nonspecific buffer association rate constant & 200 mM$^{-1}$s$^{-1}$ \\
$k_{-d}$ & \Ca-nonspecific buffer dissociation rate constant & 0.026 s$^{-1}$\\
\hline
\end{tabular}
\end{center}
\end{table*}

\begin{table*}
\begin{center}
\caption{\label{tabmb} Cytosolic material balance and calcium buffering}
\begin{tabular}{l}\hline\hline
$\D\frac{d[{\rm Na}^+]_i}{dt}  =  -\D\frac{3I_{\rm NaK}+3I_{\rm NaCa}+I_{\rm B,Na}+I_{\rm M,Na}}{F \cdot{\rm vol}_i}$ \\
$\D\frac{d[{\rm K}^+]_i}{dt}  =
-\D\frac{I_K+I_{\rm CaK}+I_{\rm Ki}+I_{\rm B,K}-2I_{\rm NaK}+I_{\rm M,K}}
                           {F \cdot{\rm vol}_i}$ \\
$\D\frac{d[{\rm Ca}^{2+}]_i}{dt} =
-\D\frac{I_{\rm Ca,L}-2I_{\rm NaCa}+I_{\rm CaP}+I_{\rm B,Ca}+I_{\rm M,Ca}+
                            I_{\rm up}-I_{\rm rel}}{2F \cdot{\rm vol}_{\rm Ca}} + \frac{d[S_{\rm CM}]}{dt}+ \frac{dB_F}{dt}$ \\
\hline
\end{tabular}
\end{center}
\end{table*}

\begin{table*}
\begin{center}
\caption{\label{tabica} L-type calcium current}
\begin{tabular}{ll}\hline\hline
$I_{Ca,L}  =  g_{\rm CaL}d_Lf_L(V_m-E_{\rm Ca,L})$ & $g_{\rm CaL} = 1.4151$ nS\\
$\D\frac{dd_L}{dt}  =  \frac{\bar{d}_L-d_L}{\tau_{d_L}}$ &
$\bar{d}_L  =  \D\frac{1.0}{1+e^{-(V_m+1.878)/7.5704}}$ \\
$\tau_{d_L}  =  2.8928 e^{-{((V_m+8.6344)/12.3884)}^2}+2.4323$ &\\
$\D\frac{df_f}{dt}  =  \frac{\bar{f}_L-f_f}{\tau_{f_f}}$ &
$\bar{f}_L  =  \D\frac{1.0}{1+e^{(V_m+29.3188)/1.5389}}$\\
$\tau_{f_f}  =  295.5937 e^{-{((V_m-4.7187)/112.545)}^2}+23.1907$ &
$f_L = 0.74f_f + 0.26$ \\
\hline
\end{tabular}
\end{center}
\end{table*}

\begin{table*}
\begin{center}
 \caption{\label{tabkca}Ca$^{2+}$-activated potassium current}
 \begin{tabular}{ll}\hline\hline
$ I_{\rm K,Ca}  =  g_{\rm KCa}P_{\rm K,Ca}(V_m-E_K)$ &\\
$ {\bar{p}_o  =  \D\frac{1.0}{1+e^{-(V_m-V_{\rm 1/2,KCa})/21.70}}}$ &
                  $ {\bar{p}_f  =  \bar{p}_s = \bar{p}_o}$ \\
$ V_{\rm 1/2,KCa}  =  -45.0\log_{10}([{\rm Ca}^{2+}]_i)-198.55$ &\\
$ \displaystyle{\frac{dp_f}{dt}  = \D
\frac{\bar{p}_f-p_f}{\tau_{\rm p_f}}}$ &
$ \displaystyle{\frac{dp_s}{dt}  = \D \frac{\bar{p}_s-p_s}{\tau_{\rm p_s}}}$ \\
$\tau_{\rm p_f} = 0.5$ ms & $\tau_{\rm p_s} = 11.5$ ms \\
$ P_{\rm KCa}  =  0.65p_f+0.35p_s $ &\\
 \hline
\end{tabular}
\end{center}
\end{table*}

\begin{table*}
\begin{center}
\caption{\label{tabik} Delayed rectifier}
\begin{tabular}{ll}\hline\hline
$I_{\rm K} = g_k{p_k}^2(V_m-E_{\rm K})$ & $g_k$=9.8325 nS \\
$\D\frac{dp_1}{dt} = \frac{\bar{p}_k-p_1}{\tau_{\rm p_1}}$ &
$\tau_{\rm p_1} = 210.9873e^{-{((V_m+214.3355)/195.3502)}^2}-20.5866$ \\
$\D\frac{dp_2}{dt} = \frac{\bar{p}_k-p_2}{\tau_{\rm p_2}}$ &
$\tau_{\rm p_2} = 821.3949e^{-{((V_m+31.5891)/27.4568)}^2}+0.1892$ \\
$\bar{p}_k = \D\frac{1.0}{1+e^{-(V_m-V_{1/2,K})/K}}$ & $V_{1/2} = -1.77$ mV, $k=14.52$ mV \\
$p_k=0.58p_1+0.42p_2$ & \\
\hline
\end{tabular}
\end{center}
\end{table*}

\begin{table*}
\begin{center}
\caption{\label{tabki} Inward rectifier}
\begin{tabular}{lr}\hline\hline
 $I_{\rm K_i} = g_{\rm max,K_i}\D\frac{V_m-E_{\rm K}}{1+e^{-(V_m-V_{1/2,K_i})/28.89}}$, & 
 $g_{\rm max,K_i} = G_{\rm K_i}{([{\rm K}^+]_o)}^{n_{\rm K_i}}$  \\
 $V_{\rm 1/2,K_i} = 25.19\log_{10}{[{\rm K}^+]_i}-112.29$ & \\
 \hline
\end{tabular}
\end{center}
\end{table*}

\begin{table*}
\begin{center}
\caption{\label{tabpex} Pump and exchanger currents}
\begin{tabular}{l}\hline\hline
$I_{\rm CaP}=\bar{I}_{\rm CaP}\D\frac{\rm CaCM}{{\rm CaCM}+K_{\rm m,CaCM}}$ \\
$I_{\rm NaK}=\bar{I}_{\rm NaK}\left(\D\frac{[{\rm K}^+]_o}{[{\rm K}^+]_o+K_{m,K}}\right)
             \left(\D\frac{[{\rm Na}^+]_i^{1.5}}{[{\rm Na}^+]_i^{1.5}+K_{\rm m,Na}^{1.5}}\right)
             \left(\D\frac{V_m+150}{V_m+200}\right)$ \\
$I_{\rm NaCa}=k_{\rm NaCa}\D\frac{[{\rm Na}^+]_i^3[{\rm Ca}^{2+}]_o\Phi_F-[{\rm Na}_+]_o^3[{\rm Ca}^{2+}]_i\Phi_R}
              {1+d_{\rm NaCa}([{\rm Na}^+]_o^3[{\rm Ca}^{2+}]_i+[{\rm Na}^+]_i^3[{\rm Ca}^{2+}]_o)}$ \\
$\Phi_F=e^{\gamma V_mF/RT}$ \ \ \ \ \ \ $\Phi_R=e^{(\gamma-1)V_mF/RT}$\\
\hline
\end{tabular}
\end{center}
\end{table*}

\begin{table*}
\begin{center}
\caption{\label{tabbk} Background currents}
\begin{tabular}{ll}\hline\hline
 $ I_b = I_{\rm b,Na}+I_{\rm b,K}+I_{\rm b,Ca}$ \\
 $ I_{\rm b,Na} = g_{\rm b,Na}(V_m-E_{\rm Na})$ & \ \ \ \ \ \ \ \ \
 $E_{\rm Na}=\D\frac{RT}{F}\ln{\D\frac{[{\rm Na}^+]_o}{[{\rm Na}^+]_i}}$ \\
 $ I_{\rm b,K} = g_{\rm b,K}(V_m-E_K)$    &  \ \ \ \ \ \ \ \ \
 $E_K=\D\frac{RT}{F}\ln{\D\frac{[{\rm K}^+]_o}{[{\rm K}^+]_i}}$ \\
 $ I_{\rm b,Ca} = g_{\rm b,Ca}(V_m-E_{\rm Ca})$ &  \ \ \ \ \ \ \ \ \
 $E_{\rm Ca}=\D\frac{RT}{2F}\ln{\D\frac{[{\rm Ca}^{2+}]_o}{[{\rm Ca}^{2+}]_i}}$ \\
  \hline
 \end{tabular}
\end{center}
\end{table*}

\begin{table*}
\begin{center}
\caption{\label{tabsr}Parameters for the multi-state ryanodine receptor (RyR) model}
\begin{tabular}{|l|l|l|}\hline
Parameter & Description & Numerical value \\\hline\hline
$K_{\rm r1}$ & activation rate constant & 2500.0 mM$^{-2}\cdot$ms$^{-1}$\\
$K_{\rm r2}$ & inactivation rate constant & 1.05 mM$^{-1}\cdot$ms$^{-1}$ \\
$K_{\rm -r1}$ & unbinding rate constant from activation & 0.0076 ms$^{-1}$\\
$K_{\rm -r2}$ & unbinding rate constant from inactivation & 0.084 ms$^{-1}$\\
$\bar{I}_{\rm up}$ & maximum SR uptaking current & 200 pA \\
$K_{\rm m,up}$ & Michaelis-Menten constant of SR calcium pump& 0.08  mM\\
$\tau_{\rm tr}$ & time constant of the internal diffusion  & 1000.0 ms \\
$\tau_{\rm rel}$ & time constant of the diffusion from SR release compartment & 0.0333 ms \\
${\rm vol}_u$ & volume of uptake compartment & 0.07 pl\\
${\rm vol}_r$ & volume of release compartment & 0.007 pl\\
$A_mP_K$ & whole-cell ${\rm K}^+$ permeability of mechanical-sensitive channel& 1.821e+06 L$\cdot$ms$^{-1}$\\
\hline
\end{tabular}
\end{center}
\end{table*}

\begin{table*}
\begin{center}
\caption{\label{tabmech} Parameters for mechanical model}
\begin{tabular}{|l|l|l|}\hline
Parameter & Description & Numerical Value \\
\hline\hline
$k_p$ & parallel element stiffness constant & 0.1 $\mu$N\\
$k_{\rm x1}$ & phosphorylated cross--bridge stiffness constant & 12.5 $\mu$N/$\mu$m\\
$k_{\rm x2}$ & latch bridge stiffness constant & 8.8 $\mu$N/$\mu$m\\
$k_s$ & series element stiffness constant & 0.2 $\mu$N\\
$\mu$ & viscosity of series viscocritical element & 0.01 $\mu$N$\cdot${ms}/$\mu$m \\
$l_0$ & cell length at zero passive force& 40 $\mu$m \\
$l_{s0}$ & length of series viscoelastic element at zero force& 30 $\mu$m \\
$l_{\rm opt}$ & optimal length of active contractile element & 100 $\mu$m \\
$f_{\rm AMp}$ & friction constant for phosphorylated cross-bridges & 1.3 $\mu$N$\cdot${ms}/{$\mu$m}\\
$f_{\rm AM}$ & friction constant for latch bridges & 85.5 {$\mu$N}$\cdot${ms}/$\mu$m\\
$v_x$ & cross-bridge cycling velocity & 5.0 $\mu$m/ms \\
$\beta$ & length modulation constant& 7.5 \\
$\alpha_p$ & constant for parallel element& 0.1 \\
$\alpha_s$ & constant for series elastic element& 4.5 \\
\hline
\end{tabular}
\end{center}
\end{table*}

\begin{table*}
\begin{center}
\caption{\label{tabpara2} \textbf{Parameters for myosin kinetics model}}
\begin{tabular}{|l|l|l|}\hline
Parameter & Description & Numerical value \\
\hline\hline
$K_2$ & myosin dephosphorylation rate constant & 0.4 s$^{-1}$\\
$K_3$ & cross--bridge formation rate constant & 1.8 s$^{-1}$\\
$K_4$ & cross--bridge detachment rate constant & 0.1 s$^{-1}$\\
$K_5$ & myosin dephosphorylation rate constant & 0.4 s$^{-1}$\\
$K_7$ & latch state to free myosin detachment rate constant & 0.045 s$^{-1}$\\
$K_{\rm CaCM}$ & half activation constant for myosin phosphorylation &  1.78e-7\\
\hline
\end{tabular}
\end{center}
\end{table*}

\begin{table*}
\begin{center}
\caption{\label{tabinit} \textbf{Initial conditions for state variables}}
\begin{tabular}{|l|l|l|}\hline\hline
$V_m=-56.1275$ mV & $p_2$=0.0159 & $R_{01}$=0.9955\\
\Cai=100.92 nM & $p_f$=0.00032 & $R_{10}$=0.0033\\
$[{\rm Na}^+]_i$ = 5.06 mM & $p_s$=0.00032 & $R_{11}$=4.0e-06\\
$[{\rm K}^+]_i$ = 91.03 mM & $[S_{\rm CM}]$=0.072 mM & $[{\rm Ca}^{2+}]_{\rm up}$ = 0.5687 mM\\
$d_L$ = 0.00046 & $[B_F]$=0.336 mM& $[{\rm Ca}^{2+}]_r$=0.5503 mM\\
$f_f$ = 1.000 & M = 0.6099 & $l_s$=37.05 $\mu$m\\
$f_s$ = 1.000& Mp = 0.0476& $l_x$=89.60 $\mu$m\\
$p_1$ = 0.0159& AMp = 0.0627 & \\
\hline
\end{tabular}
\end{center}
\end{table*}

\end{document}